\documentclass[useAMS,usenatbib]{mnras}
\usepackage{graphicx}
\title[IFU spectroscopy of southern PNe: VII]
  {IFU spectroscopy of Southern PNe: VII Photo-ionization modelling of intermediate excitation class objects}
\author[Ali \& Dopita.]
  {A. Ali,$^{1,2}$ M.A. Dopita,$^{3}$   \\
  $^1$Department of Astronomy, Faculty of Science, Cairo University, 12613 Giza, Egypt \\
  $^2$Astronomy Dept, Faculty of Science, King Abdulaziz University, 21589 Jeddah, Saudi Arabia \\
  $^3$Research School of Astronomy and Astrophysics, Australian National University, Cotter Rd., Weston ACT 2611, Australia
  }
\date{Released 2018}

\pagerange{\pageref{firstpage}--\pageref{lastpage}} \pubyear{2018}

\def\LaTeX{L\kern-.36em\raise.3ex\hbox{a}\kern-.15em
    T\kern-.1667em\lower.7ex\hbox{E}\kern-.125emX}

\begin{document}

\label{firstpage}

\maketitle

\begin{abstract}

We present integral field unit spectroscopic observations of southern Galactic planetary nebulae (PNe), IC\,2501, Hen\,2-7, and PB\,4.
The goal of studying these objects together is that, although they have roughly similar intermediate excitation and evolution of central stars (CSs), they display very different evolution in their nebular structure which needs to be understood.  The morphologies and ionisation structures of the objects are investigated using a set of emission-line maps representative of the different ionisation zones. We use those in order to construct two-zone self-consistent photoionisation models for each nebula to determine new model-dependent distances, progenitor luminosities, effective temperatures and CS masses. The physical conditions, chemical compositions, and expansion velocities and ages of these nebulae are derived. In Hen\,2-7 we discover a strong poleward-directed jet from the presumed binary CS. Oxygen and nitrogen abundances derived from both collisionally excited and recombination lines reveal that PB\,4 displays an extreme abundance discrepancy factor, and we present evidence that this is caused by fluorescent pumping of the O\,II ion by the EUV continuum of an interacting binary CS, rather than by recombination of the O\,III ion. Both IC\,2501 and PB\,4 were classified by others as Weak Emission Line Stars (WELS). However, our emission line maps show that their recombination lines are spatially extended in both objects, and are therefore of nebular rather than CS origin. Given that we have found this result in a number of other PNe, this result casts further doubt on the reliability, or even the reality, of the WELS classification.

\end{abstract}


\section{Introduction}

Photoionisation modelling of planetary nebulae is vital to an understanding of their CSs and their evolution. On one hand, such models enable us to derive reliable abundances especially for elements observed in only one ionisation zone or those which are observed only in weak emission lines, such as the recombination lines of heavy elements. On other hand, such models enable us to determine the nebular temperature and density in the different nebular ionisation zones. Furthermore, they can provide information on the distance and evolutionary age of the nebula and are necessary to determine the temperature, luminosity, mass, and radius of the CSs. The location of CS on the luminosity-temperature plane overlaid by the H-burning or He-burning post AGB tracks such as those of \citet{VW94} and/or \citet{Miller16}, provide us with the mass of the precursor star, so that the chemical composition can be related directly to the nature of the precursor star. Reviewing the literature,  photoionisation modelling of PNe has been carried out by, for example, \citet{Henry15, Bohigas15, Bohigas13}, \citet{Yuan11}, \citet{Pottasch11, Pottasch09}, \citet{Morisset09},  \citet{Pottasch05}, \citet{Surendiranath08}, \citet{Surendiranath04}, \citet{Hyung04}, and \citet{Ercolano03} but up to now relatively few of these are based on integral field spectroscopy of the full nebula.

Although \citet{Weidmann11}  identified $\sim13$\% of CSs in a sample of slightly over 3000 PNe, the fundamental difficulty in observing the central stars is that they are weak in visible light since most of their radiation is in the ultraviolet range. Furthermore, the observed spectra of these stars are often contaminated with nebular emission which may lead to errors in their classification. Therefore, high resolution spectra integral field unit (IFU) are important to be able to properly subtract the contaminating nebular mission and to correctly classify the PN central star.

IFU spectroscopy of PNe also provides an excellent opportunity for constructing accurate nebular models. In particular, we can directly compare the global spectrum to a theoretical model using the size, nebular flux and morphology to constrain the distance, the photoionisation structure, and to define the inner and outer boundaries of the nebula.  In addition, we can directly compare measured electron temperatures and densities in the different ionisation zones to constrain the pressure distribution within the ionised material.

In this series of papers, our group has provided IFU images, spectra and detailed PNe photoionisation models for a number of southern PNe. To briefly summarise the key results; self-consistent photoionisation and shock models were constructed to interpret the physics of the interacting planetary nebula  PNG342.0-01.7 \citep{Ali15b}. \citet{Basurah16} introduced self-consistent modelling for a sample of high-excitation non-Type I PNe with supposed weak emission-line central stars, showing these lines to arise in the nebula rather than the central star. A detailed self-consistent model was developed for the compact, young, and low excitation class PN IC\,418 using high spectral resolution IFU spectra covering the spectral range 3300-8950\AA\, \citep{Dopita17}. This model consists of three separate zones: an inner photoionised shock driven by the accelerating stellar wind of the CS, a photoionised nebular shell, and an outer shock in the AGB wind driven by the over-pressure of the strong D-Type ionisation front.  Very recently, \citet{Dopita18} have developed a set of self-consistent radiative shock models to investigate the physical conditions and peculiar chemo-dynamics of the N-rich fast-moving knots associated with the bipolar PN Hen 2-111.

In this article, we continue our program of obtaining IFU spectra with both high resolution and very high dynamic range and building detailed photoionisation models. Here we study IC\,2501, Hen\,2-7, and PB\,4. These PNe were chosen to be studied together because, although having both very similar excitation classes and luminosities of the central stars, they display remarkably heterogenous morphologies, possibly related to the binarity or otherwise of their central stars.

A long-slit spectrum of IC\,2501 has been obtained by \citet{Milingo02a} in the spectral range 3600-9600\,\AA\, with wavelength dispersions of 2.8\AA. The plasma diagnostics and abundances determination were given in a companion paper \citep{Milingo02b}. An echelle spectrum was obtained by \citet{Sharpee07} to identify emission lines of the s-process elements Br, Kr, Xe, Rb, Ba, and Pb in this nebula. Their spectrum cover the spectral ranges 3280-4700\,\AA\, and 4590-7580\,\AA\, at very high resolution ($\lambda/\Delta \lambda = 28,000$ and 22,000, respectively).  Also, a few krypton lines at [Kr III]\,6826\AA, [Kr IV]\,5346\AA, and [Kr IV]\,5868\AA\,were detected in the spectrum of IC\,2501 by \citet{Sterling15}. In addition to the emission lines in optical band, infrared features at 17.4 and 18.9\,$\micron$ of C$_{60}$ were detected in IC\,2501 \citep{Otsuka14}.  \citet{Cuisinier96} were obtained long-slit spectra for Hen\,2-7 and PB\,4 in the wavelength range 3600-7400\AA\, with a spectral resolution of $\sim$ 4.0\,\AA. Based on these observations, they derived the physical conditions, ionic abundances and total elemental abundances of both nebulae. H$\alpha$+[N II] and [O III] narrow-band images of IC\,2501 and PB\,4  were presented by \citet{Schwarz92}. Both images  reveal circular an elliptical morphologies for IC\,2501 and PB\,4, respectively.  Another set of H$\alpha$+[N II] and [O III] images for PB\,4 were presented by \citet{Corradi03} which show a double PN halo. The inner halo is brighter and structured in H$\alpha$+[N II] while the outer halo is more asymmetrical and fragmentary. A narrow band image in [N\,II] filter, with bandpass filter 18\,\AA, was given by \citet{Weidmann16} for Hen\,2-7. The image shows ``a well-defined elliptical nebula with a high surface brightness and a prominent central star".

\citet{Peimbert78} originally divided PNe into four types according to their chemical composition, spatial and kinematic characteristics. Type I objects are helium and nitrogen rich, Type II intermediate population, Type III high velocity, and Type IV halo population. Type I objects are those satisfy the condition He/H $\geqslant$ 0.14 or N/O $\geqslant$ 1.0. Subsequently, \citet{Peimbert83} relaxed the Type I condition to include all objects with He/H $\geqslant$ 0.125 or N/O $\geqslant$ 0.5. Later, \citet{Maciel99}  considered a more strict condition, defining Type I as those objects have He/H $>$ 0.125 and N/O $>$ 0.5.   Members of Type II are characterised by He/H $<$ 0.125 and N/O $<$ 0.5. \citet{Faundez87} further subdivided Type II PNe into two types according to their nitrogen abundance. \citet{Quireza07} re-analysed the Peimbert classes through a statistical study of a large sample of PNe to remove the confusion concerning the objects which cannot be defined as belonging to a single type.  They define the limits between the four Peimbert types on the basis of helium and nitrogen abundances, nitrogen to oxygen ratio, the height above the Galactic disk, and the peculiar velocity of each object (see Table 2, \citet{Quireza07}).

\begin{table*}
\centering
\caption{The observing log}
\label{Table1}
\scalebox{0.95}{
\begin{tabular}{llccccc}
\hline
Object (PNG number) & No. of   & Exposure  & Date   & Airmass & Standard \& telluric stars\\
	    &            frames   & time (s)  &        &         &               \\
\hline \hline
IC\,2501 (PN G281.0-05.7)   \\
B7000 \& I7000 &     8        & 10(2), 40(2), 150(2), 600(2) & 04/04/2016  & 1.19 & HD 111980 \& HIP 41423\\
U7000 \& R7000 &     8        & 10(2), 40(2), 150(2), 600(2) & 04/04/2016  & 1.19 & HD 111980 \& HIP 41423\\
\\
Hen\,2-7 (PN G264.2-08.1)  \\
B7000 \& I7000 &    9  & 10(3), 60(3), 900(3)    & 12/01/2016 &  1.10 & HD 031128 \& HIP 41423 \\
U7000 \& R7000 &    9  & 10(3), 100(3), 1000(3)  & 12/01/2016 &  1.10 & HD 031128  \& HIP 41423 \\
\\
PB\,4 (PN G275.0-04.1)     \\  	
B7000 \& I7000 &    5 & 50(3), 1200(2)    & 06/04/2016 &  1.09 & HD111980 \& HIP 41423  \\	
U7000 \& R7000 &    5 & 100(3), 1200(2)   & 06/04/2016 &  1.09 & HD111980 \& HIP 41423  \\	
\hline
\end{tabular}}
\end{table*}

The main objective of this paper is to provide IFU spectra  for the Galactic PNe IC\,2501, Hen\,2-7, and PB\,4  at high resolution ($R \sim 7000$), to present a detailed spectroscopic and morphological study and to construct self-consistent photoionisation models for these objects. This  paper  is  structured  as  follows.  The observations and data reduction are explained in Section  2.  Section 3 provides analysis of the plasma diagnosis while Section 4 is devoted for describing the morphologies, expansion velocities and distances of the sample in addition to the binarity of the PB\,4 CS. Section 5 discusses the misclassification of IC\,2501 and PB\,4 central stars as weak emission-line stars type. The photoionisation modelling of IC\,2501, Hen\,2-7, and PB\,4 are presented in Section 6 and the conclusions are given in Section 7.

\section{OBSERVATIONS \& DATA REDUCTION}

\begin{table*}
\centering
\caption{Integrated $F(\lambda)$ and de-reddened $F_d(\lambda)$ line fluxes (relative to H$\beta = 100$) for IC\,2501, Hen\,2-7, and PB\,4. The full version of the table is available  online. Only a fraction of the table is presented here to draw the attention of the reader to its content.}
\label{Table2}
\scalebox{0.90}{
\begin{tabular}{llllcllcll}
\hline
$\lambda(\rm \AA)$	& Ion  &  \multicolumn{2}{c}{IC 2501}&& \multicolumn{2}{c}{Hen 2-7} && \multicolumn{2}{c}{PB 4} \\\cline{3-4}\cline{6-7}\cline{9-10}
	\\
                &       & $F(\lambda)$  & $F_d(\lambda)$  && $F(\lambda)$  & $F_d(\lambda)$ && $F(\lambda)$  & $F_d(\lambda)$\\
\hline													
3721.63	&	[S~\textsc{iii}]	&	2.26	$\pm$	0.23	&	3.07	$^{+	0.32		}_{	-0.36	}$	&	&					&					&	&	1.41	$\pm$	0.33	&	2.03	$^{	+0.464}_{	-0.492}$	\\
3726.03	&	[O~\textsc{ii}]	&	35.40	$\pm$	1.19	&	46.50				$\pm$	2.60		&	&		39.08	$\pm$	0.808	&	52.3	$\pm$	3.3		&	&	7.02	$\pm$	0.83	&	10.0	$^{	+1.200}_{	-1.400}$	\\
3728.82	&	[O~\textsc{ii}]	&	15.47	$\pm$	0.63	&	20.50				$\pm$	1.20		&	&		30.19	$\pm$	1.162	&	40.8	$\pm$	2.9		&	&	3.44	$\pm$	0.48	&	4.91	$^{	+0.681}_{	-0.790}$	\\
3734.37	&	H~\textsc{i}	&	2.14	$\pm$	0.21	&	2.63	$^{+	0.30		}_{	-0.31	}$	&	&		1.42	$\pm$	0.117	&	1.91	$^{	+0.186}_{	-0.206}$	&	&	1.3	$\pm$	0.38	&	1.87	$^{	+0.546}_{	-0.547}$	\\
3749.49	&	O~\textsc{ii}	&				&								&	&		1.92	$\pm$	0.112	&	2.67	$^{	+0.211}_{	-0.229}$	&	&				&					\\
3750.15	&	H~\textsc{i}	&	2.43	$\pm$	0.25	&	3.31	$^{+	0.35		}_{	-0.39	}$	&	&					&					&	&	2	$\pm$	0.44	&	2.85	$^{	+0.627}_{	-0.669}$	\\
3759.87	&	O~\textsc{iii}	&				&								&	&		0.18	$\pm$	0.01	&	0.23	$^{	+0.018}_{	-0.020}$	&	&	1.31	$\pm$	0.29	&	1.87	$^{	+0.409}_{	-0.424}$	\\
3770.63	&	H~\textsc{i}	&	2.82	$\pm$	0.29	&	3.99	$^{+	0.41		}_{	-0.45	}$	&	&		2.83	$\pm$	0.148	&	3.72	$^{	+0.285}_{	-0.308}$	&	&	2.54	$\pm$	0.57	&	3.62	$^{	+0.821}_{	-0.822}$	\\
3797.9	&	H~\textsc{i}	&	3.76	$\pm$	0.39	&	5.37	$^{+	0.54		}_{	-0.60	}$	&	&		4.1	$\pm$	0.2	&	5.19	$^{	+0.385}_{	-0.415}$	&	&	3.42	$\pm$	0.67	&	4.85	$^{	+0.965}_{	-0.957}$	\\
3819.62	&	He~\textsc{i}	&	1.18	$\pm$	0.08	&	1.44	$^{+	0.11		}_{	-0.12	}$	&	&		1.15	$\pm$	0.052	&	1.41	$^{	+0.101}_{	-0.109}$	&	&	0.84	$\pm$	0.25	&	1.17	$^{	+0.351}_{	-0.346}$	\\
3835.39	&	H~\textsc{i}	&	5.07	$\pm$	0.34	&	6.54	$^{+	0.50		}_{	-0.54	}$	&	&		5.43	$\pm$	0.088	&	7.12	$\pm$	0.415		&	&	4.72	$\pm$	0.78	&	6.62	$^{	+1.130}_{	-1.140}$	\\
3868.75	&	[Ne~\textsc{iii}]	&	59.65	$\pm$	14.81	&	76.00	$^{+	19.10		}_{	-19.60	}$	&	&		79.95	$\pm$	1.312	&	105	$\pm$	6		&	&	56.15	$\pm$	4.57	&	77.7	$^{	+6.800}_{	-7.500}$	\\													

\hline
\end{tabular}}
\end{table*}

The IFU spectroscopic observations of Hen\,2-7, IC\,2501, and PB\,4  were acquired at 12 January, 4 April, and 6 April 2016, respectively, and cover the wavelength range 3300-8950\,\AA.  The observations were obtained with single pointings using the Wide Field Spectrograph (WiFeS) instrument \citep{Dopita07, Dopita10} mounted on the 2.3-m ANU telescope at Siding Spring Observatory. This instrument delivers a field of view of 25 $\times$ 38 at a spatial resolution of either 1.0 $\times$ 0.5 or 1.0 $\times$ 1.0, depending on the binning on the CCD. By using a series of exposures stepped in integration times,  a very high dynamic range $(\sim 10^{5-6})$ can be achieved.  Also, the high resolution R$\sim7000$ gratings were employed, providing a full-width-half-maximum resolution of $\sim 45$ km/s ($\sim 0.9$\,\AA). Observations are made simultaneously in two gratings. For the U7000 \& R7000 gratings, the RT480 dichroic was used, which cuts at 480nm and for the B7000 \& I7000 gratings, the RT615 was employed, which cuts at 615nm. Therefore each waveband is observed in a region of high dichroic efficiency. A suitably wide overlap in wavelength coverage is ensured between each of the gratings \citep{Dopita07}, giving a contiguous wavelength coverage from $\sim 3300$ to $\sim 8950$\AA. A summary of the WiFeS observations is presented in Table \ref{Table1}.

The data reduction was carried out using the PyWiFeS pipeline \citep{Childress14}. The nebular fluxes were calibrated using the STIS spectrophotometric standard stars HD 111980 and HD 031128. The wavelength scale was calibrated using observations of the Ne-Ar arc Lamp throughout the night. Furthermore, a telluric standard star HIP 41423 was used to improve the removal of the OH and H$_2$O telluric absorption features in the red. The separation of these features by molecular species allows for a more accurate telluric correction by accounting for night to night variations in the column density of these two species. For further details on this process, please refer to \citet{Childress14}.

Each of these nebulae are fairly compact and fall well within the WiFeS field of view,  having angular diameters (in their largest dimension) of 10\,arc sec. (IC2501), 17\,arc sec. (Hen 2-7) and  12.5\,arc sec. (PB 4). The global spectra of each PN were extracted from the reduced data cubes utilising a circular aperture matched to the observed size of the PN using {\tt QFitsView v3.1} software\footnote{a FITS file viewer using the QT widget library developed at the Max Planck Institute for Extraterrestrial Physics by Thomas Ott}. The spectra from the four gratings U, B, R, and I were combined applying the {\tt scombine} task of the {\tt  IRAF} software.

\section {Plasma diagnosis}
\subsection{Line fluxes and excitation class}

Emission-line fluxes and their uncertainties were measured from the final combined flux-calibrated U, B, R, and I spectra using the {\tt  ALFA} code \citep{Wesson16}. The uncertainties are estimated using the noise structure of the residuals.  A double check was done using the {\tt splot} task in the IRAF software.

The computation of the interstellar reddening coefficients and the subsequent plasma diagnoses steps used the Nebular Empirical Abundance Tool ({\tt NEAT}; \citet{Wesson12}). The line fluxes were treated for the reddening effect applying the extinction law of \citet{Howarth83}. The reddening coefficient ${c(H\beta)}$ was determined from the weighted mean ratios of the Hydrogen Balmer lines H$\alpha$, H$\beta$, H$\gamma$, and H$\delta$ in an iterative method, assuming Case B at $T_e = 10^4$\,K.  The observed and de-reddened line fluxes and their uncertainties are given in Table \ref{Table2}.

The He\,II $\lambda$4686/H$\beta$ line ratio probably provides the best estimator for the nebular excitation class (EC). Here, we applied the  scheme of \citet{Reid10} to determine the EC of  IC\,2501, Hen\,2-7, and PB\,4. The He\,II$\lambda$4686 line is marginally present in both IC\,2501 and Hen\,2-7. All three of these PNe are of intermediate EC class (EC $\sim 6$). This result is compatible with the absence of high excitation lines, such as [Ar\,V] and [Ne\,V], in the spectra.

The systemic velocity RV$_{\rm sys}$ of the sample was determined using the IRAF external package RVSAO (emsao task), from numerous nebular emission lines. The heliocentric radial velocity RV$_{\rm hel}$ was calculated by correcting the RV$_{\rm sys}$ for the effect of Earth's motion.  The results of IC\,2501 and Hen\,2-7 reveal good agreement with those of \citet{Durand98}. It appears that there are no previous measurements for the radial velocity of PB\,4.

Table \ref{Table3} lists the H$\alpha$ and H$\beta$ integrated fluxes, ${c(H\beta)}$, EC, RV$_{\rm hel}$ and distance (Section \ref{Distance}) of each PN. Almost all measurements are well consistent with those in the literature.
\\
\begin{table*}
	\centering \caption{Reddening coefficient, observed $H\beta$ flux, observed
		$H\alpha$ flux, excitation class, heliocentric radial velocity and distance of IC 2501, Hen 2-7, and PB 4.}
	
	\label{Table3} \scalebox{0.90}{
		\begin{tabular}{lcccccccccccc}
			\hline
			Object & \multicolumn{2}{c}{$c(H\beta)$} & \multicolumn{2}{c}{Log F$(H\beta)$}  & \multicolumn{2}{c}{Log F$(H\alpha)$} & \multicolumn{1}{c}{EC} & \multicolumn{2}{c}{RV$_{\rm hel}$ (km/s)}  & \multicolumn{3}{c}{Distance (kpc)}\\
			& Obs. & Others  & Obs. & Others & Obs. & Others &   & Obs. & (6) & (Dist 1) & (Dist 2) & adopted\\
			\hline
			IC 2501   & 0.546    & 0.41$^1$, 0.59$^2$ , 0.56$^3$ &-10.63 & -10.70$^3$, -10.67$^4$ & -10.00 & -10.01$^2$ & 6.2 & 26.9$\pm$2.8 & 31.5$\pm$0.2 & 2.40 & 2.88 & 2.64\\			
			Hen 2-7  & 0.496     & 0.53$^1$, 0.39$^2$, 0.63$^5$ & -11.43
			 & -11.40$^4$, -11.85$^5$ & -10.82 & -10.81$^2$ & 6.1 & 85.8$\pm$2.1& 88.0$\pm$4.0 & 2.90 & 3.29 & 3.10\\
			PB 4 & 0.738  &  0.76$^1$, 0.53$^2$, 0.60$^5$  & -11.70
			  &  -11.66$^4$, -12.15$^5$ & -11.00 & -11.03$^2$ &  6.8 & 92.3$\pm$2.6 & & 3.14 & $3.0^{*}$ & 3.05\\
			\hline
\end{tabular}}
\begin{minipage}[!t]{17cm}
		{{\bf References}: 			
			(1) \citet{Tylenda92};(2) \citet{Frew13}; (3) \citet{Milingo02a}; (4) \citet{Cahn92}; (5) \citet{Cuisinier96}}; (6) \citet{Durand98}. \\
{\bf Dist 1 \& Dist 2} are derived following \citet{Ali15a} and \citet{Frew16} distance scales. \\
		$^{*}$ This value is calculated following the \citet{Frew16} distance equation for optically thin planetary nebulae.\\
\end{minipage}
\end{table*}
\\

\subsection{Temperatures and densities from CELs}

The electron temperatures and densities for IC\,2501, Hen\,2-7, and PB\,4 are calculated from their collisional excitation lines (CELs) using the  NEAT code. The emission lines which are detected in the PNe spectra allow us to measure both electron temperatures and densities for several stages of ionisation. The nebular temperatures are determined from the line ratios   [O III] ($\lambda$4959 + $\lambda$5007)/$\lambda$4363, [Ar\,III] ($\lambda$7135 + $\lambda$7751)/$\lambda$5192, [N\,II] ($\lambda$6548 + $\lambda$6584)/$\lambda$5754, [S II] ($\lambda$6717 + $\lambda$6731)/($\lambda$4068+ $\lambda$4076), [O II] ($\lambda$7319 + $\lambda$7330)/($\lambda$3726+ $\lambda$3729), and [O\,I] ($\lambda$6363 + $\lambda$6300)/$\lambda$5577 while nebular densities were determined from the line ratios [O\,II] $\lambda$3727/$\lambda$3729, [S\,II] $\lambda$6716/$\lambda$6731, [Cl\,III] $\lambda$5517/$\lambda$5537, and [Ar IV] $\lambda$4711/$\lambda$4740. In Table \ref{Table4}, we list the temperatures and densities of IC\,2501, Hen\,2-7, and PB\,4, and compare these values  with those available in the literature and with those derived from their photoionisation models presented in Section 6, below. There is generally good agreement between the literature values and those presented here.

\begin{table*}
\caption{Electron temperatures and densities of IC\,2501, Hen\,2-7, and PB\,4 compared with other works.}
\label{Table4}
\scalebox{0.65}{
		\begin{tabular}{lllllllllll}
			\hline
Object   & \multicolumn{6}{|c|}{Temperature (K) from CELs} && \multicolumn{3}{|c|}{Temperature (K) from ORLs}\\
\cline{2-7} \cline{9-11}
\\
& {[}O III{]} & {[}Ar III{]} & {[}N II{]} &  {[}S II{]} & {[}O II{]} & {[}O I{]} && He I & O II &  H\,I Paschen jump \\
\hline
IC\,2501 (This work) & 9350$^{+390}_{-390}$ & 9250$^{+250}_{-250}$ & 11200$^{+300}_{-300}$ & 12600$^{+4500}_{-2600}$ & 11100$^{+3340}_{-2340}$ & 9140$^{+570}_{-450}$ && 9275$^{+2212}_{-2275}$ & 5748$^{+1431}_{-1282}$ &  10622 \\
IC\,2501 (Model) & 11260 & 11270 & 11110 & 10440 & 11230 & 9742 \\
\citet{Milingo02b} & 9500 &  & 11200 & 11700 & 10600  \\
\citet{Sharpee07} & 9500$^{+300}_{-200}$ & 9400$^{+600}_{-500}$ & 10800$^{+900}_{-1100}$ & 12000 & 13000 & 6900$^{+300}_{-200}$ \\
\hline

Hen\,2-7 (This work) & 13100$^{+300}_{-300}$ & 10700$^{+400}_{-400}$ & 11600$^{+200}_{-200}$ & & 13409$^{+670}_{-670}$ & 9918$^{+1190}_{-1190}$ &&  6950$^{+500}_{-500}$ & & 7017 \\
Hen\,2-7 (Model) & 12604 & 12408 & 12068 & 11803 & 12676 & 9885 \\
\citet{Cuisinier96} & 11700 &  & 11800 \\
\hline

PB\,4$^a$ (This work) & 10000$^{+500}_{-500}$ &  & 14600$^{+3000}_{-2800}$ &  & $ > 19000^b$ &&  & 3221$^{+579}_{-965}$ &  & 7637 \\
 &  &  & 9500$^{+1000}_{-1000}$$^d$ & - & 17800$^{+2000}_{-2000}$$^d$  &  &  & &  \\
PB\,4 (Model) & 11050 & 11020 & 11022 & 11010 & 11016 & 11009 \\
\citet{Cuisinier96} & 9400  \\
\hline
Object   &  \multicolumn{4}{|c|}{Density (cm$^{-3}$) from CELs} && \multicolumn{3}{|c|}{Density (cm$^{-3}$) from ORLs}\\  \cline{2-5} \cline{7-9}
\\
 &  {[}O II{]} & {[}S II{]} & {[}Cl III{]} & {[}Ar IV{]} && O\,II & N\,II & H\,I Paschen decrement  \\
\hline

IC 2501 (This work) & 7580$^{+2590}_{-1690}$  & 8830$^{+2140}_{-1600}$  & 8380$^{+480}_{-450}$ & 10700$^{+2800}_{-2200}$ && 3980$^c$ & 1270$^c$ & 7874 \\
IC 2501 (Model) &  8563 & 8401 & 8553 & 8486 \\
\citet{Milingo02b} &   & 4800  &  &  \\
\citet{Sharpee07} &  11000$^{+9000}_{-4000}$  &11000  & 8500$^{+2100}_{-1600}$ & 8775$^{+2150}_{-1600}$ \\
\hline

Hen 2-7 (This work) &  1160$^{+160}_{-140}$ &   881$^{+66}_{-61}$  & 1010$^{+130}_{-110}$ & 1600$^{+370}_{-360}$ && 1584$^c$ &  & 7997$^{b}$  \\
Hen 2-7 (Model) &  808 &  712 & 807 & 774 \\
\citet{Cuisinier96} &  &  1500  &  &  \\
\hline

PB 4$^a$ (This work) &  3091$^{+1220}_{-676}$ &   2632$^{+220}_{-200}$  & 1230:$^{+710}_{-620}$ & $ 1916^{+1510}_{-1230}$ && 7080$^c$ & 2300$^c$ & 1061 \\
PB 4 (Model) & 2169 & 1853 & 2126 \\
\citet{Cuisinier96} &  &  3970  &  & \\
\hline
\end{tabular}}
\begin{minipage} [!t]{17cm}
$^a$ PB 4 is defined as optically thin nebula following the criteria proposed by \citet{Kaler89}. The PN spectra shows the absence of [O I] and [N I] lines, weak flux of low-excitation lines such as [O II] and [N II], and the very weak flux of [N II] relative to  H$\alpha$ ($\rm {[N II]/H}\alpha < 0.1$).\\
$^{b}$ Probably unreliable; $^{c}$ The uncertainty is of order $10-30\%$; $^{d}$ The value after correction for recombination contribution.\\
\end{minipage}
\end{table*}

\subsection{Temperatures and densities from ORLs}
The PNe spectra declare few optical recombination lines (ORLs), convenient for electron temperature and density diagnostics. The temperatures were derived from the diagnostic ratios of He I $λ5876/λ4471$, $λ6678/λ4471$, $λ6678/λ5876$, and $λ7281/λ5876$, and  O\,II  $λ4649/λ4591$, $λ4649/λ4189$, and $λ4649/λ4089$. Further, the densities were derived from the diagnostic ratios of O\,II $λ4649/λ4662$ and $λ4076/λ4070$ and N\,II $λ5679/λ5666$. The average temperature and density that determined from each ion were listed in Table \ref{Table4}, providing the diagnostic lines are available in the nebular spectra.

Jointly with the temperatures and densities derived from both CELs and ORLs, Table \ref{Table4} also gives the Paschen jump temperature and the Paschen decrement density applying the NEAT code. We ignored here the Balmer jump temperature and the Balmer decrement density due to the S/N is too low in the UV spectral region of the three nebulae.

\citet{Rubin86} has examined the effects of the recombination processes in addition to the collisional excitations on the energy level populations of species e.g. nitrogen and oxygen. The recombination contributions of N$^{2+}$ and O$^{2+}$ in the strength of the auroral [N\,II] $\lambda5754$ and the [O\,II] $\lambda\lambda 7320, 7330$ lines were estimated following Equations 1 and 3 of \citet{Liu00}, respectively.
For the [N\,II] weak line $\lambda5754$, we estimate recombination contribution of  $4\%$ and $70\%$ of the observed intensity of $\lambda5754$ in IC\,2501 and PB\,4, respectively. For the [O\,II] line $\lambda\lambda 7320, 7330$, we estimate recombination contribution of $4\%$ and $65\%$ in the observed intensity of $\lambda\lambda 7320, 7330$ in IC\,2501 and PB\,4, respectively. It is apparent the high contribution of the recombination in the strength of lines $\lambda5754$ and $\lambda\lambda 7320, 7330$ in case of PB\,4 compared to IC\,2501. This result has a significant effect on the temperatures of PB\,4 which derived from both [N\,II] and [O\,II] line ratios. Subtracting the recombination contribution, we obtain a temperatures of 9500K from corrected [N\,II] ($\lambda$6548 + $\lambda$6584)/$\lambda$5754 line ratio and 17600k from corrected [O\,II] ($\lambda$7319 + $\lambda$7330)/($\lambda$3726+ $\lambda$3729) line ratio.

\begin{table}
	\centering \caption{Fractional ionic abundances of $\rm O^{2+}$ and $\rm N^{2+}$ lines in IC\,2501 and PB\,4.} \label{Table5}
\scalebox{0.7} {\begin{tabular}{l@{\hspace{0.0cm}}l@{\hspace{0.1cm}}ll@{\hspace{0.0cm}}l@{\hspace{0.1cm}}l}		
			\hline \\
			$\lambda(\rm \AA)$	&	Multiplet	&	$\frac{X(line)}{H^{+}}$  & $\lambda(\rm \AA)$	&	Multiplet	&	$\frac{X(line)}{H^{+}}$ \\
			\hline \\
\multicolumn{6}{c}{IC\,2501} \\ \cline{1-6}
\\
  \multicolumn{3}{l}{$\rm O^{2+}/H^{+}$} &   \multicolumn{3}{l}{$\rm N^{2+}/H^{+}$}\\
 \\
		4649.13 & V1& 4.54E-4($\pm$2.60E-5)             & 5666.63  & V3 & 1.97E-4($\pm$2.6E-5) \\
        4661.63 & V1 & 6.34E-4($\pm$2.90E-5)            & 5676.02  & V3 & 2.33E-4($\pm$3.60E-5) \\
                &\bf{V1} & \bf{5.03E-04($\pm$2.68E-05)}   & 5679.56  & V3 & 1.99E-04($\pm$1.00E-05) \\
      4414.90 & V5 & 5.74E-04($\pm$1.78E-04)              & 5710.77  & V3 & 4.01E-04($\pm$8.50E-05) \\
      4416.97 & V5 & 1.01E-03($\pm$2.90E-04)              &      & \bf{V3}& \bf{2.27E-04($\pm$2.67E-05)} \\
      4452.37 & V5 & 2.33E-03($\pm$5.00E-04)              & 4630.54  & V5 & 3.41E-04($\pm$3.20E-05) \\
              &\bf{V5} & \bf{1.08E-03($\pm$2.84E-04)}     &      & \bf{V5}& \bf{3.41E-04($\pm$3.20E-05)} \\
      4072.16 & V10 & 5.84E-04($\pm$5.80E-05)             & 4788.13  & V20 & 2.27E-04($\pm$5.40E-05) \\
              & \bf{V10} & \bf{5.84E-04($\pm$5.80E-05)}   &    & \bf{V20}& \bf{2.27E-04($\pm$5.40E-05)} \\
      4132.80 & V19 & 6.57E-04($\pm$6.60E-05)             & {$\rm N^{2+}/H^{+}$} & &\bf{2.65E-04($\pm$1.32E-04)} \\
      4153.30  & V19 & 7.71E-04($\pm$7.80E-05)  & \\
               & \bf{V19} & \bf{7.28E-04($\pm$7.35E-05)}    & \\
      4890.86  & V28 & 1.99E-03($\pm$4.10E-04)  & \\
      4906.83 & V28 & 1.15E-03($\pm$1.70E-04)  & \\
              & \bf{V28} & \bf{1.53E-03($\pm$3.02E-04)}  & \\
     4089.29  & V48a & 4.58E-04($\pm$4.60E-05)  & \\
              & \bf{V48a} & \bf{4.58E-04($\pm$4.60E-05)}  & \\
 {$\rm O^{2+}/H^{+}$} && \bf{5.68E-04($\pm$5.11E-05)}  & \\
\\
\\
\multicolumn{6}{c}{PB\,4} \\ \cline{1-6}
\\	
		4641.81  & V1   & 5.30E-03($\pm$5.80E-04)  & 5666.63  & V3 & 1.49E-03($\pm$1.30E-04) \\
        4649.13  & V1   & 4.41E-03($\pm$5.20E-04)  & 5676.02  & V3 & 1.44E-03($\pm$1.30E-04) \\
        4661.63  & V1   & 5.10E-03($\pm$7.30E-04)  & 5679.56  & V3 & 1.48E-03($\pm$1.00E-04) \\
      & \bf{V1}  & \bf{4.81E-03($\pm$5.73E-05)}    & 5686.21  & V3 & 1.52E-03($\pm$2.50E-04) \\
        4069.62  & V10  & 9.00E-03($\pm$2.00E-03)  & 5710.77  & V3 & 2.50E-03($\pm$1.29E-04) \\
         4072.16 & V10  & 6.00E-03($\pm$1.00E-03)  &          & \bf{V3} & \bf{1.58E-03($\pm$1.29E-04)} \\
         4075.86 & V10  & 4.34E-03($\pm$8.50E-04)  & 5931.78  & V28 & 7.94E-04($\pm$7.70E-05) \\
      & \bf{V10} & \bf{6.06E-03($\pm$9.20E-04)}    & 5941.65  & V28 & 5.99E-04($\pm$4.30E-05) \\
        4924.53  & V28  & 5.00E-03($\pm$1.00E-03)  & 5952.39  & V28 & 1.57E-03($\pm$2.10E-04) \\
      & \bf{V28} & \bf{5.00E-03($\pm$1.00E-03)}    &          & \bf{V28} & \bf{8.69E-04($\pm$1.47E-04)} \\
        4609.44  & V92a & 4.00E-03($\pm$1.50E-03)  &   {$\rm N^{2+}/H^{+}$} & & \bf{1.22E-03($\pm$1.38E-04)} \\
      & \bf{V92a}& \bf{4.00E-03 ($\pm$1.50E-03)}   &           &         & \\
       {$\rm O^{2+}/H^{+}$} && \bf{4.97E-03($\pm$6.40E-04)}  &  &  &  \\	

\hline \\
	\end{tabular}}
\end{table}

\subsection{Ionic and elemental abundances}
\begin{table*}
	\centering \caption{Ionic and total abundances of IC\,2501, Hen\,2-7, and PB\,4.} \label{Table6}
\scalebox{0.75}{
		\begin{tabular}{llll}
			
			\hline
			\\
			Ions        &   IC\,2501 &  Hen\,2-7 & PB\,4 \\
			\hline
			\\
{\bf CEL abundances} \\														
N$^{+}$/H	&	1.15E-05	(	1.60E-06	)	(	-1.40E-06	)	&	9.70E-06	(	1.15E-06	)	(	-1.03E-06	)	&	7.14E-07	(	4.90E-08	)	(	-4.60E-08	)	\\
ICF(N)	&	2.32E+01	(	5.56E-05	)	(	-4.40E-05	)	&	9.89E+00	(	8.48E-01	)	(	-7.81E-01	)	&	6.26E+01	(	1.20E+01	)	(	-1.01E+01	)	\\
N$^{}$/H	&	2.66E-04	(	5.50E-05	)	(	-4.60E-05	)	&	9.59E-05	(	1.66E-05	)	(	-1.41E-05	)	&	4.48E-05	(	9.30E-06	)	(	-7.70E-06	)	\\
N$^{}$/H (Model) & 7.41E-05  & 9.3E-05 & 3.23E-05 \\
\\																									
O$^{0}$/H	&	8.80E-06	(	2.00E-06	)	(	-1.87E-06	)	&	5.72E-06	(	1.37E-06	)	(	-1.56E-06	)	&	4.12E-06	(	4.37E-07	)	(	-4.40E-07	)							\\
O$^{+}$/H	&	5.33E-05	(	9.60E-06	)	(	-7.90E-06	)	&	2.46E-05	(	1.30E-06	)	(	-1.20E-06	)	&	4.54E-06	(	3.20E-07	)	(	-3.00E-07	)	\\
O$^{2+}$/H	&	4.44E-04	(	6.90E-05	)	(	-6.00E-05	)	&	1.63E-04	(	1.60E-05	)	(	-1.50E-05	)	&	2.71E-04	(	5.30E-05	)	(	-4.10E-05	)	\\
ICF(O)	&	1.00E+00	(	0.00E+00	)	(	0.00E+00	)	&	1.00E+00	(	1.00E-02	)	(	-1.00E-02	)	&	1.00E+00	(	1.00E-02	)	(	-1.00E-02	)	\\
O$^{}$/H	&	4.98E-04	(	7.00E-05	)	(	-6.20E-05	)	&	1.93E-04	(	1.70E-05	)	(	-1.60E-05	)&	2.79E-04	(	5.90E-05	)	(	-4.50E-05	)	\\
O$^{}$/H (Model) & 2.34E-04   & 2.00E-04 & 2.34E-04\\
\\																									
Ne$^{2+}$/H	&	1.11E-04	(	2.80E-05	)	(	-2.40E-05	)	&	5.25E-05	(	3.00E-06	)	(	-2.80E-06	)	&	8.27E-05	(	2.06E-05	)	(	-1.44E-05	)	\\
ICF(Ne)	&	1.22E+00	(	3.00E-02	)	(	-2.50E-02	)	&	1.19E+00	(	4.30E-02	)	(	-3.40E-02	)	&	1.12E+00	(	9.00E-03	)	(	-9.00E-03	)	\\
Ne$^{}$/H	&	1.35E-04	(	3.30E-05	)	(	-2.90E-05	)	&	6.28E-05	(	4.10E-06	)	(	-3.80E-06	)	&	9.25E-05	(	2.28E-05	)	(	-1.61E-05	)	\\
Ne$^{}$/H (Model) & 5.37E-05  & 4.68E-05 & 5.89E-05\\
\\																									
Ar$^{2+}$/H	&	1.74E-06	(	2.20E-07	)	(	-1.90E-07	)	&	8.33E-07	(	1.24E-07	)	(	-1.08E-07	)	&	1.07E-06	(	1.70E-07	)	(	-1.40E-07	)	\\
Ar$^{3+}$/H	&	2.10E-07	(	2.80E-08	)	(	-2.50E-08	)	&	3.00E-07	(	2.10E-08	)	(	-2.00E-08	)	&	7.35E-07	(	1.48E-07	)	(	-1.23E-07	)	\\
ICF(Ar)	&	1.27E+00	(	3.80E-02	)	(	-4.10E-02	)	&	1.02E+00	(	3.20E-02	)	(	-1.70E-02	)	&	1.13E+00	(	4.70E-02	)	(	-4.70E-02	)	\\
Ar$^{}$/H	&	2.47E-06	(	3.50E-07	)	(	-3.10E-07	)	&	1.16E-06	(	1.70E-07	)	(	-1.50E-07	)	&	2.04E-06	(	3.30E-07	)	(	-2.80E-07	)	\\
Ar$^{}$/H (Model) & 1.00E-06  & 1.10E-06 & 1.15E-06 \\
\\																									
S$^{+}$/H	&	3.72E-07	(	6.50E-08	)	(	-4.90E-08	)	&	4.10E-07	(	5.00E-08	)	(	-4.40E-08	)	&	4.36E-08	(	2.60E-09	)	(	-2.40E-09	)	\\
S$^{2+}$/H	&	3.62E-06	(	9.90E-07	)	(	-8.50E-07	)	&	--							&	1.45E-06	(	4.20E-07	)	(	-3.20E-07	)	\\
ICF(S)	&	1.48E+00	(	9.60E-02	)	(	-9.30E-02	)	&	1.54E+01	(	1.20E+00	)	(	-1.10E+00	)	&	2.72E+00	(	1.57E-01	)	(	-1.20E-01	)	\\
S$^{}$/H	&	5.84E-06	(	1.74E-06	)	(	-1.34E-06	)	&	6.33E-06	(	1.06E-06	)	(	-9.10E-07	)	&	4.07E-06	(	1.37E-06	)	(	-1.02E-06	)	\\
S$^{}$/H (Model) & 5.01E-06  & 7.08E-06 & 4.57E-6\\
\\																									
Cl$^{+}$/H	&	6.48E-09	(	8.80E-10	)	(	-7.70E-10	)	&	4.90E-09	(	8.50E-10	)	(	-7.20E-10	)	&								\\
Cl$^{2+}$/H	&	8.07E-08	(	9.60E-09	)	(	-8.50E-09	)	&	4.40E-08	(	4.40E-09	)	(	-4.00E-09	)	&	8.52E-08	(	1.58E-08	)	(	-1.33E-08	)	\\
Cl$^{3+}$/H	&	8.82E-09	(	1.25E-09	)	(	-1.09E-09	)	&	2.20E-08	(	4.00E-09	)	(	-3.40E-09	)	&	4.29E-08	(	7.00E-09	)	(	-6.00E-09	)	\\
ICF(Cl)	&	1.00E+00	(	0.00E+00	)	(	0.00E+00	)	&	1.00E+00	(	0.00E+00	)	(	0.00E+00	)	&	1.00E+00	(	0.00E+00	)	(	0.00E+00	)	\\
Cl$^{}$/H	&	9.61E-08	(	1.11E-08	)	(	-9.90E-09	)	&	7.10E-08	(	8.90E-09	)	(	-7.90E-09	)	&	1.28E-07	(	2.20E-08	)	(	-1.90E-08	)	\\
Cl$^{}$/H (Model) & 6.76E-08  & 6.31E-08 & 1.38E-07 \\
\\
N/O     &   0.53 & 0.50 & 0.15 \\
N/O (Model) & 0.32 & 0.50 & 0.14 \\

{\bf RL abundances} \\
He$^{+}$/H	&	1.03E-01	(	6.00E-03	)	(	-6.00E-03	)	&	1.01E-01	(	5.00E-03	)	(	-5.00E-03	)	&	1.12E-01	(	5.00E-03	)	(	-5.00E-03	)	\\
He$^{2+}$/H	&	1.82E-05	(	5.60E-06	)	(	-5.50E-06	)	&	1.71E-03	(	8.00E-05	)	(	-8.00E-05	)	&	2.40E-02	(	2.00E-03	)	(	-2.00E-03	)	\\
He/H	&	1.03E-01	(	6.00E-03	)	(	-6.00E-03	)	&	1.03E-01	(	5.00E-03	)	(	-5.00E-03	)	&	1.36E-01	(	6.00E-03	)	(	-6.00E-03	)	\\
He/H (Model) & 1.15E-01  & 1.05E-01 & 1.44E-01\\
\\																									
C$^{2+}$/H	&	9.38E-04	(	5.10E-05	)	(	-5.10E-05	)	&	1.77E-04	(	1.50E-05	)	(	-1.40E-05	)	&	5.64E-03	(	8.20E-04	)	(	-8.30E-04	)	\\
C$^{3+}$/H	&	2.26E-05	(	1.70E-06	)	(	-1.60E-06	)	&	6.10E-05	(	6.70E-06	)	(	-6.10E-06	)	&	1.51E-04	(	2.00E-05	)	(	-1.80E-05	)	\\
ICF(C)	&	1.00E+00	(	0.00E+00	)	(	0.00E+00	)	&	1.00E+00	(	0.00E+00	)	(	0.00E+00	)	&	1.00E+00	(	0.00E+00	)	(	0.00E+00	)	\\
C/H	&	9.60E-04	(	5.30E-05	)	(	-5.00E-05	)	&	2.38E-04	(	1.70E-05	)	(	-1.50E-05	)	&	5.79E-03	(	8.40E-04	)	(	-8.40E-04	)	\\
\\
N$^{2+}$/H	&	2.65E-04	(	1.32E-4	)	(	-1.32E-4	)	&		&	1.22E-03	(	1.38E-04	)	(	-1.38E-04	)	\\
ICF(N)	    &	1.00E+00	(	0.00E+00)	(	0.00E+00	)	&		&	1.00E+00	(	0.00E+00	)	(	0.00E+00	)	\\
N$^{}$/H	&	8.14E-04	(	1.32E-4	)	(	-1.32E-4	)	&		&	1.22E-03	(	1.38E-04	)	(	-1.38E-04	)	\\
adf(N)      &   1.00E+00                                        &       &   27.2E+00 \\	
\\																							
O$^{2+}$/H	&	5.68E-04	(	5.11E-05)	(	-5.11E-05	)	&		&	4.97E-03	(	6.40E-04	)	(	-6.40E-04	)	\\
ICF(O)	    &	1.00E+00	(	0.00E+00)	(	0.00E+00	)	&		&	1.00E+00	(	1.00E-02	)	(	-1.00E-02	)	\\
O$^{}$/H	&	5.68E-04	(	5.11E-05)	(	-5.11E-05	)	&		&	4.97E-03	(	6.40E-04	)	(	-6.40E-04	)	\\
adf($\rm O^{2+}$)      &   1.14E+00                                        &       &   18.30E+00 \\

\hline			
	\end{tabular}}
\end{table*}

In Table \ref{Table6} we present the ionic and elemental abundances of IC\,2501, Hen\,2-7, and PB\,4 as derived using the {\tt NEAT} code. The ionic abundances of nitrogen, oxygen, neon, argon, sulphur, and chlorine are derived from the CELs, while helium and carbon are calculated from the optical recombination lines (ORLs) using the appropriate temperature and density for their ionisation zone. When several lines are observed for the same ion the average abundance was adopted. The total elemental abundances were determined from the ionic abundances applying the ionisation correction factors (ICFs) given by \citet{Delgado14}. Following the Peimbert classification scheme of PNe as modified by \citet{Quireza07}, none of the three nebulae studied here are of Type I. We classify IC\,2501 as of Type IIa, Hen\,2-7 as Type IIb, and PB\,4 as Type IIb/III.

The oxygen and nitrogen ionic and total abundances in IC\,2501 and PB\,4 are also computed using the ORLs. The O\,II abundance of IC\,2501 was determined from V1, V5, V10, V19, V28, and V48 multiplets, which mostly agree well with each other excepting multiplets V5 and V28 which give higher values. We  used the remaining multiplets (V1, V10, V19, and V48) to compute the fractional and overall oxygen abundance. The N\,II abundance of IC\,2501 was determined from V3, V5, and V20 multiplets.  We ignored a few blended lines in these multiplets, which gave higher abundance compared to other components. The abundance of each  multiplet was computed from a flux weighted average of its  components and the overall abundance are determined as the average of the multiplet abundances.

In the case of PB\,4, we calculate the O\,II abundance from V1, V10, V28, and V92 multiplets, which display very good agreement with each other, and the N\,II abundance from V3 and V28 multiplets.  Table \ref{Table5} lists the fractional ionic and overall abundances of O\,II and N\,II along with the average of each multiplet. The number of O\,II and N\,II lines detected in Hen\,2-7 is not sufficient to calculate their abundances.

The $\rm O^{2+}$ adf is defined as the ratio of ORL abundance of $\rm O^{2+}$ to CEL of $\rm O^{2+}$ and N adf as the ratio of ORL abundance of $\rm N^{2+}$ to the total CEL nitrogen abundance. The resultant abundance discrepancy factors (adfs) are given in Table \ref{Table5}, for both oxygen and nitrogen. No abundance discrepancy appears  in IC\,2501, where the oxygen and nitrogen abundances determined from both CELs and ORLs are roughly equal; adf($\rm O^{2+}$) $\sim$ 1.1 and adf(N) $\sim$ 1.0.  On the contrary, PB\,4 displays an extreme abundance discrepancy;  adf($\rm O^{2+} \sim 18$) and adf(N)$\sim 27$.

The derived adf of oxygen in case of PB\,4 is to be more trusted than that of nitrogen. The reason for this is that the ORL nitrogen abundance is determined mainly from $\rm N^{2+}$ abundance while the nitrogen abundance from the CELs is determined only from $\rm N^{+}$ lines and relies heavily on the estimated on the ICF . This becomes particularly uncertain for CELs where a small amount of nitrogen ($\sim$ 16\%) is in the form of $\rm N^{+}$. This result of high O adf joins PB\,4 to the group of PNe with extreme adfs (adf $>$ 10, \citet{Wesson18}).

\section{Emission line maps and expansion velocities}

\subsection{Emission line maps}\label{maps}

\begin{figure*}
	{ \begin{tabular}{@{}ccc@{}}
			\includegraphics[scale=0.60]{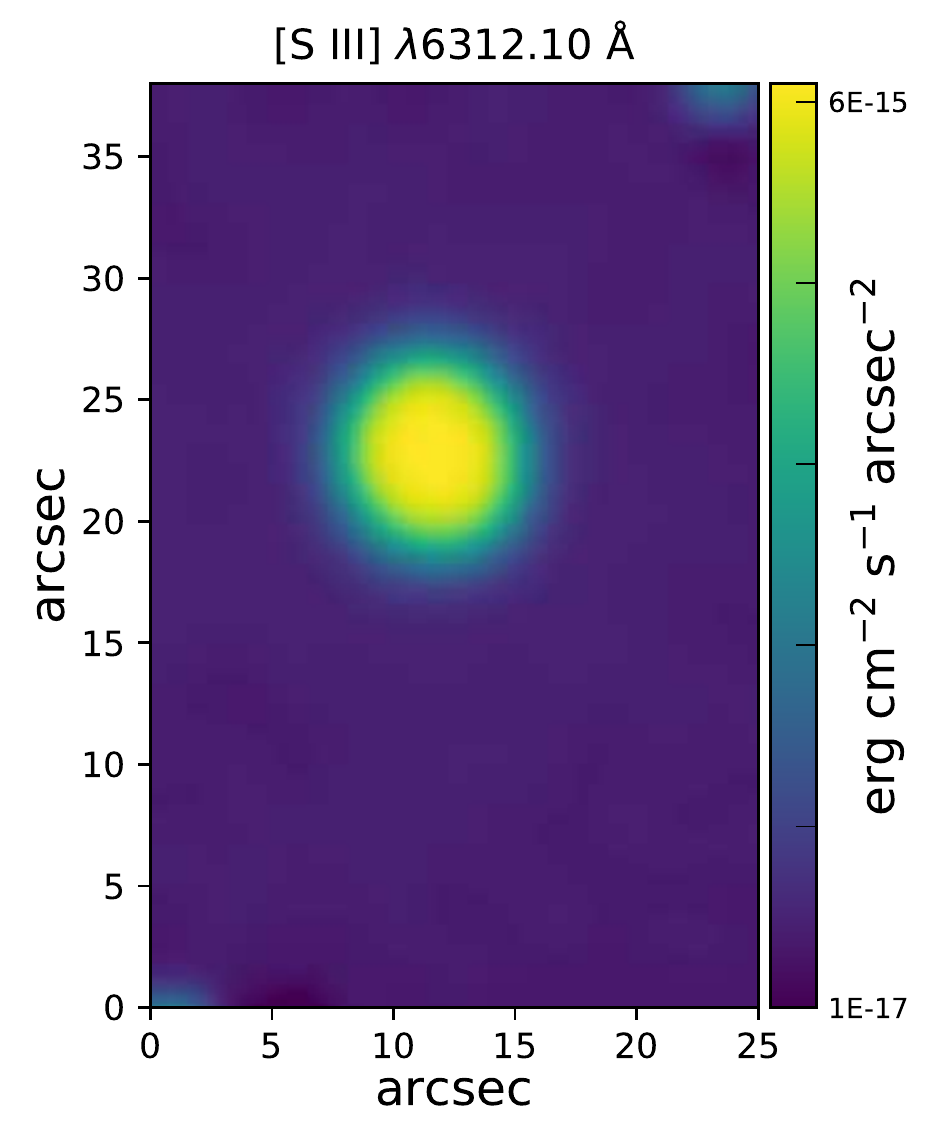}&
			\includegraphics[scale=0.60]{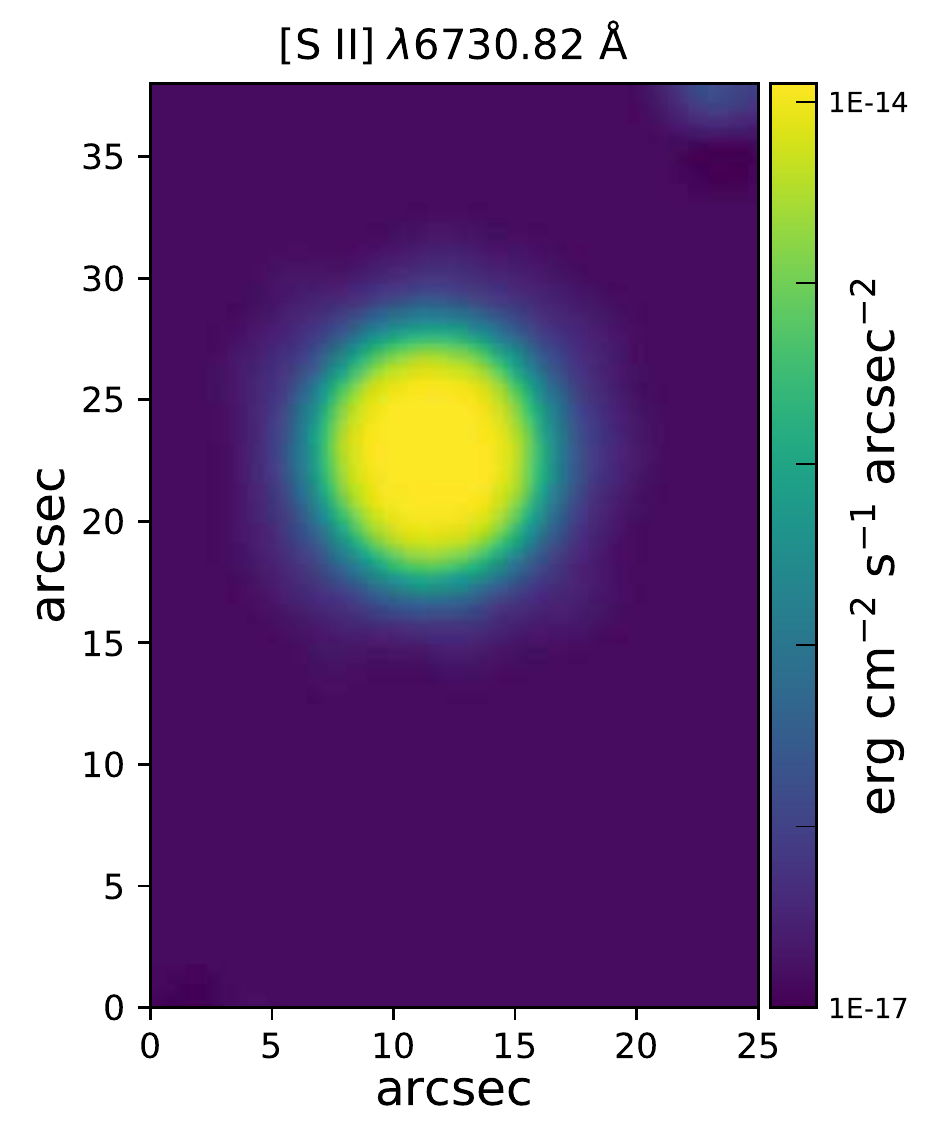}&
			\includegraphics[scale=0.60]{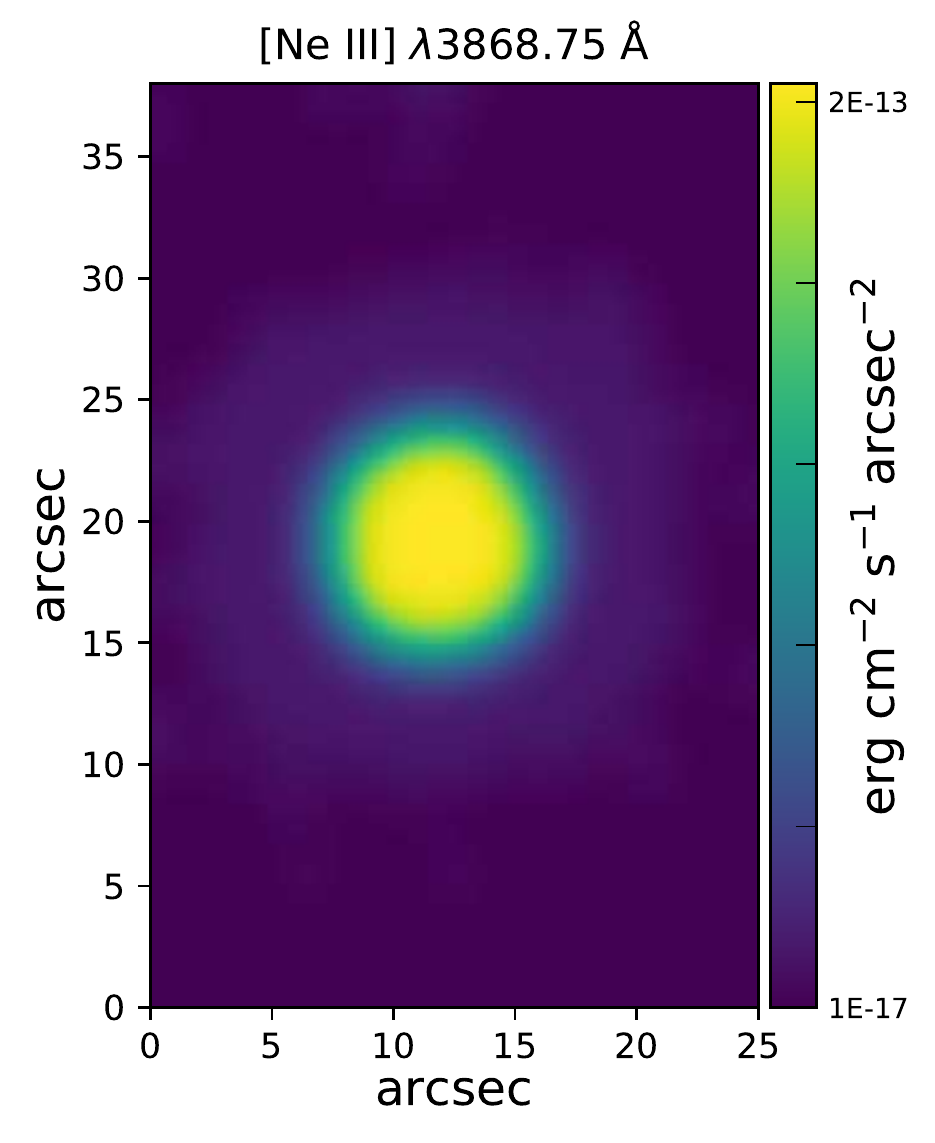}
	\end{tabular}}
	\caption{The emission-line maps of IC\,2501 in three different ions: [S\,III] at 6312\,\AA (left panel), [S\,II] at 6730\,\AA (middle panel), and [Ne\,III] at 3868\,\AA (right panel). The rough elliptical morphology are seen in all maps, with indication to double shells clearly seen in [Ne\,III] and [S\,II] images. The emission of the outer shell is slightly higher than the background sky emission. The North is to the top of the image and East is to the left.} \label{Figure_1}
\end{figure*}

\begin{figure*}
	{ \begin{tabular}{@{}ccc@{}}
			\includegraphics[scale=0.60]{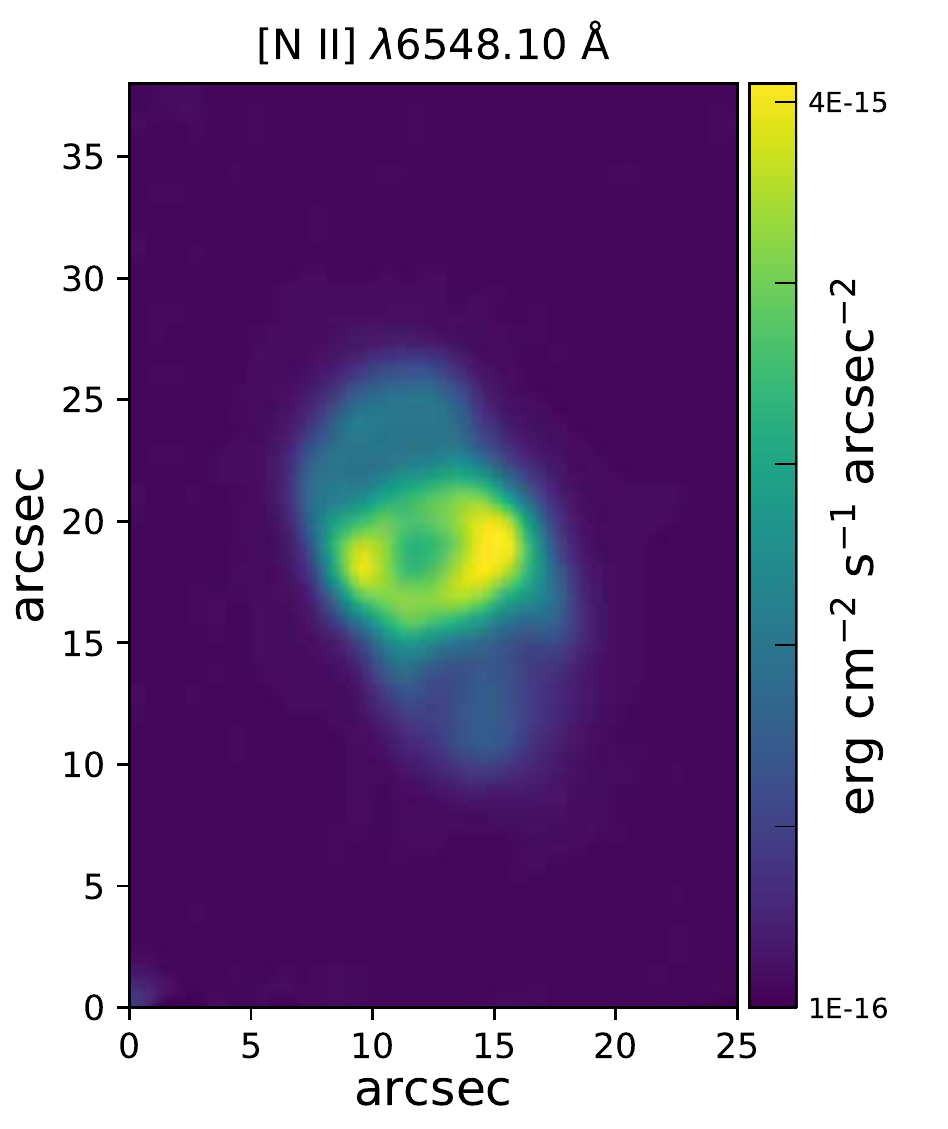} &
			\includegraphics[scale=0.60]{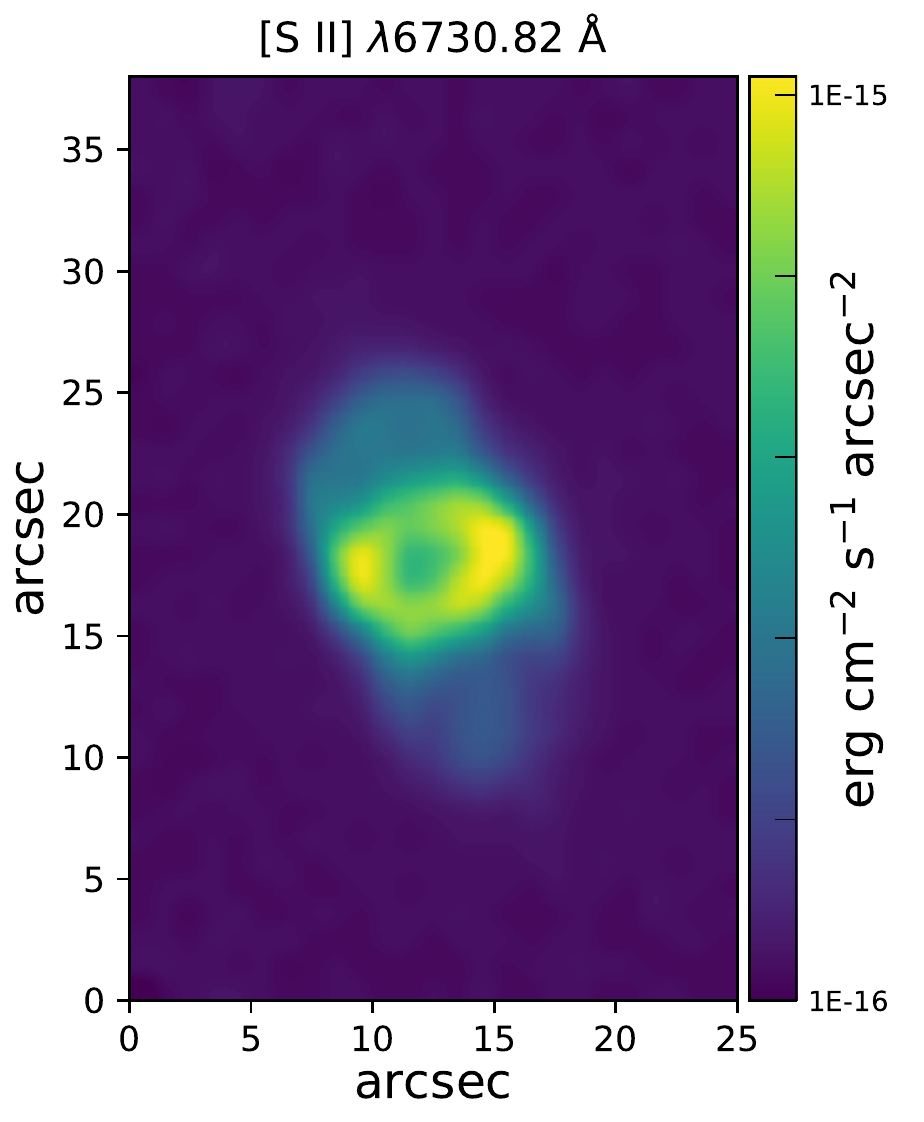} &
			\includegraphics[scale=0.60]{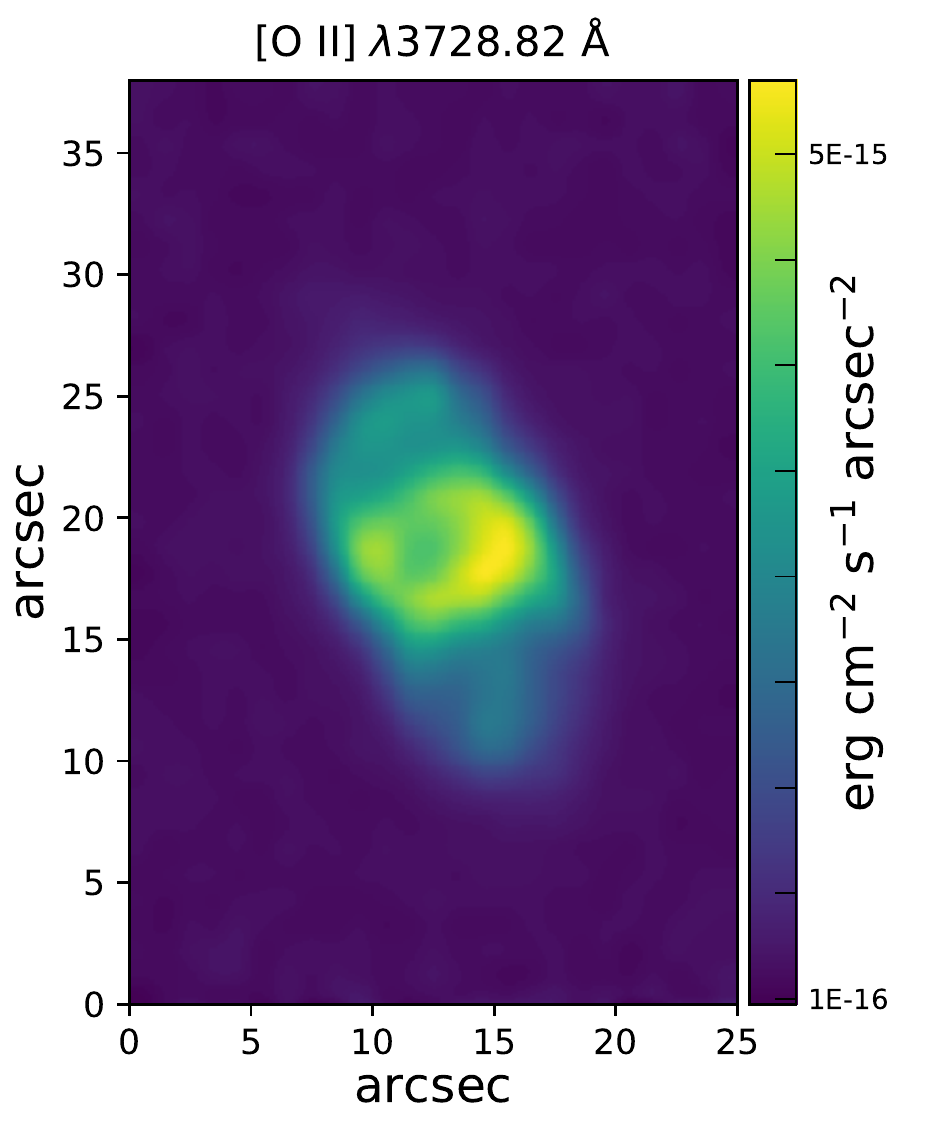}\\
			\includegraphics[scale=0.60]{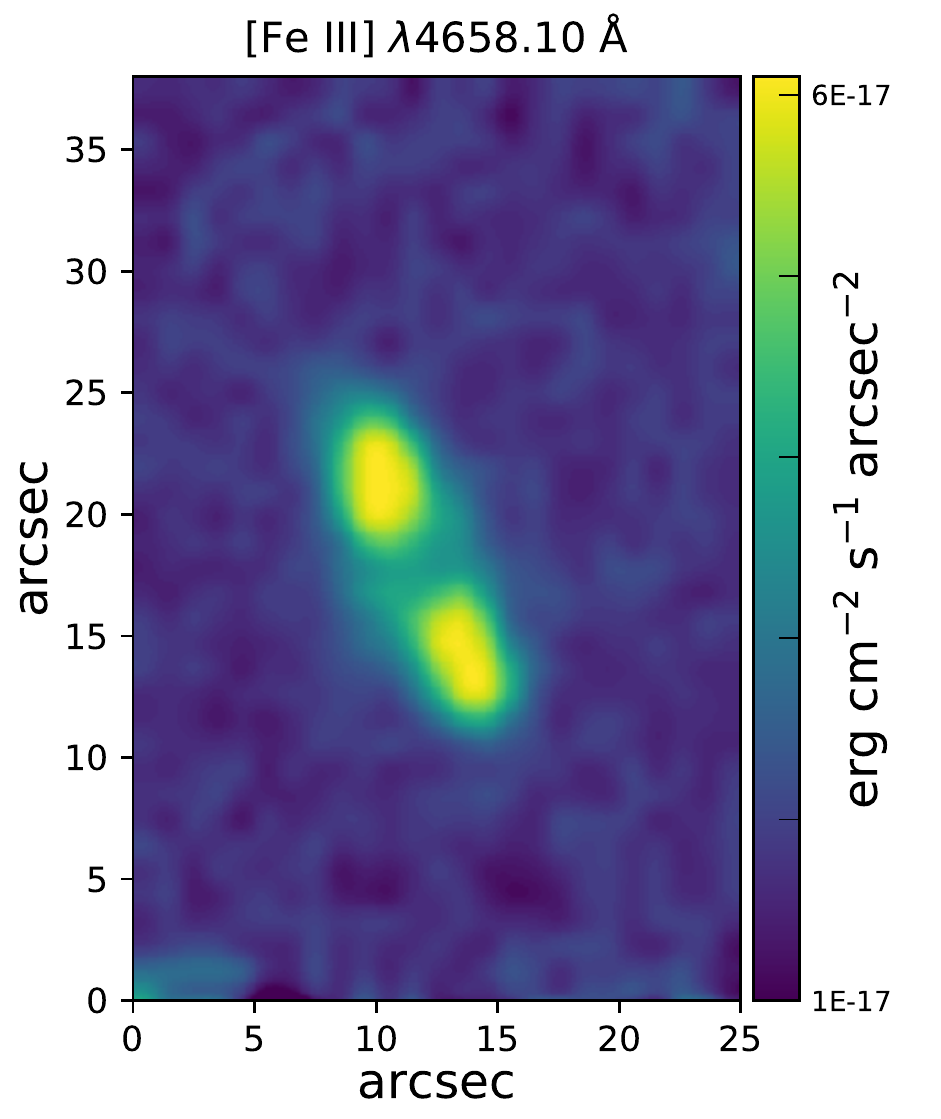} &
			\includegraphics[scale=0.60]{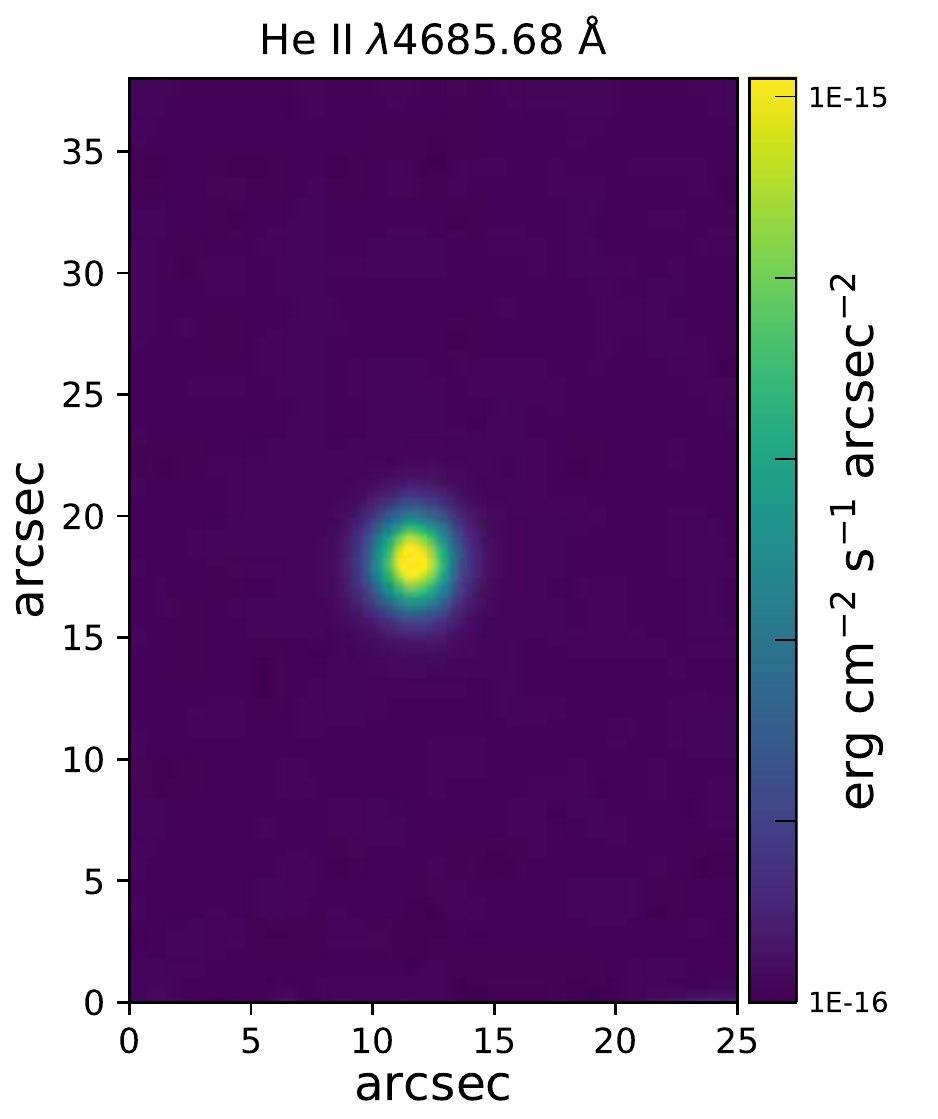} &
			\includegraphics[scale=0.60]{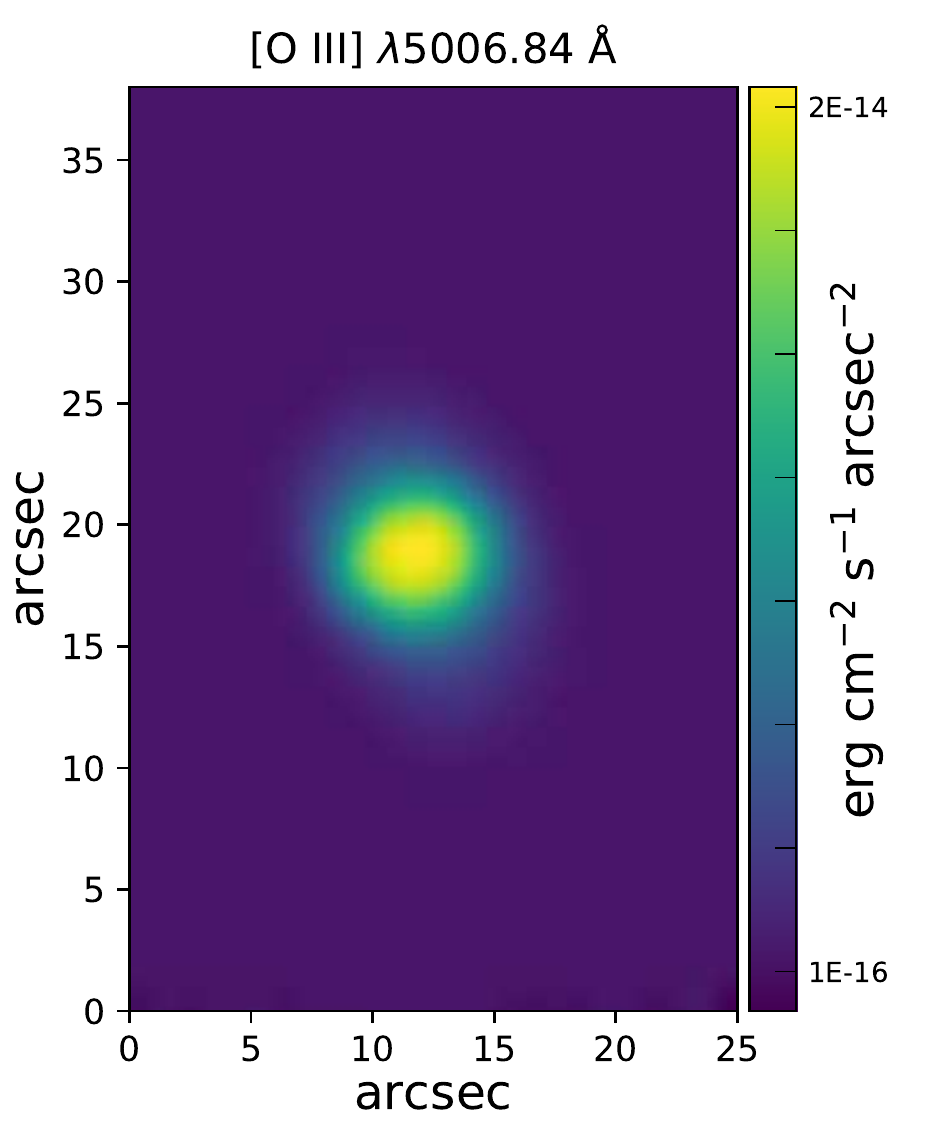}
	\end{tabular}}
	\caption{The emission-line maps of Hen\,2-7 nebula in six different ions: [N\,II] at 6548\,\AA (upper left panel), [S\,II] at 6730\,\AA (upper middle panel), and [O\,II] at 3729\,\AA (upper right panel), [Fe\,III] at 4658\,\AA (lower left panel), [He\,II] at 4686\,\AA (lower middle panel), and [O\,III] at 5007\,\AA (lower right panel). In this figure and subsequent figures, North is at the right of the image and East is to the top. A butterfly morphology appears in the low ionisation lines of nitrogen, sulphur, and oxygen. A pronounced dense tours of gas appears mainly in yellow colour mixed with less emission in green colour. By  contrast with the butterfly morphology, Hen\,2-7 has an elliptical appearance in both the He\,II and [O\,III] emission lines, while the [Fe\,III] emission-line map shows a jet-like morphology with obvious two symmetric condensations of gas along the major (polar) axis.} \label{Figure2}
\end{figure*}

\begin{figure*}
	{ \begin{tabular}{@{}ccc@{}}
			\includegraphics[scale=0.60]{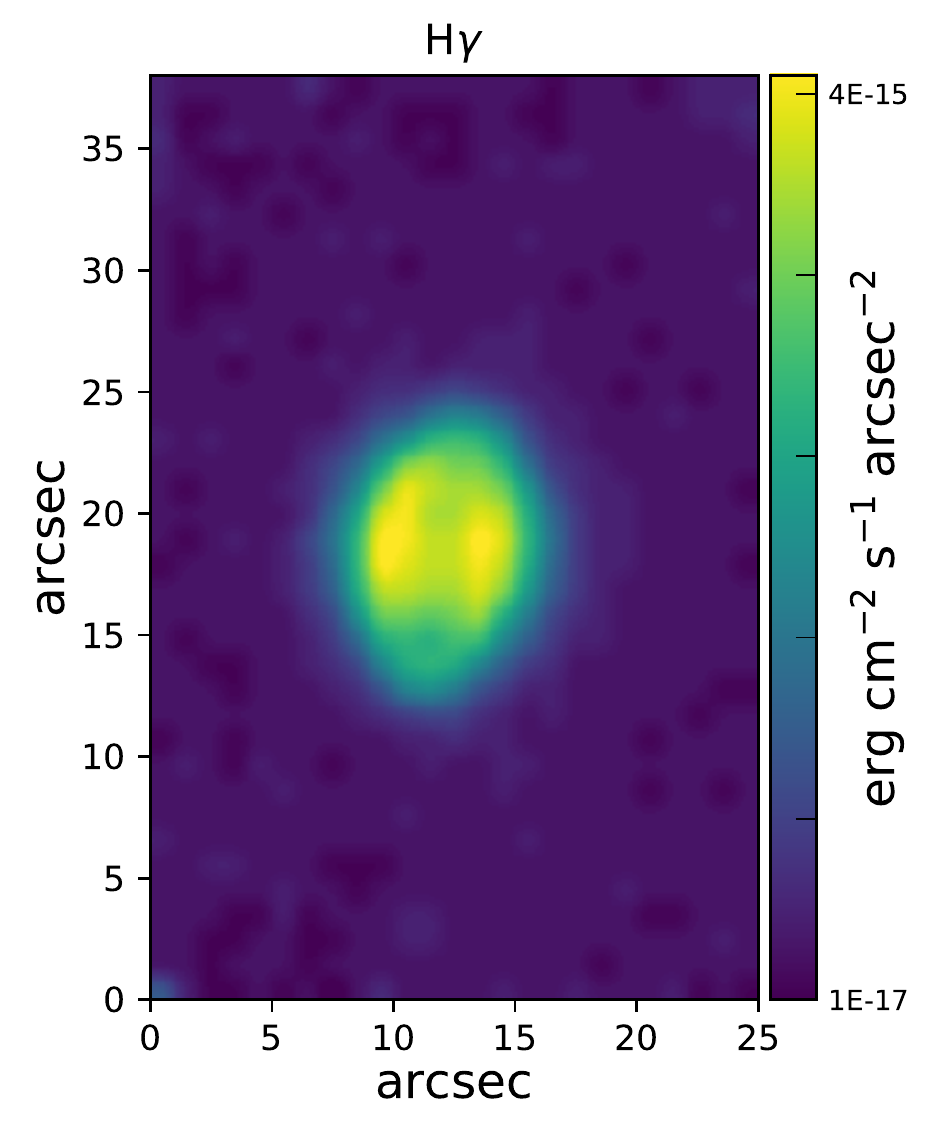}&
			\includegraphics[scale=0.60]{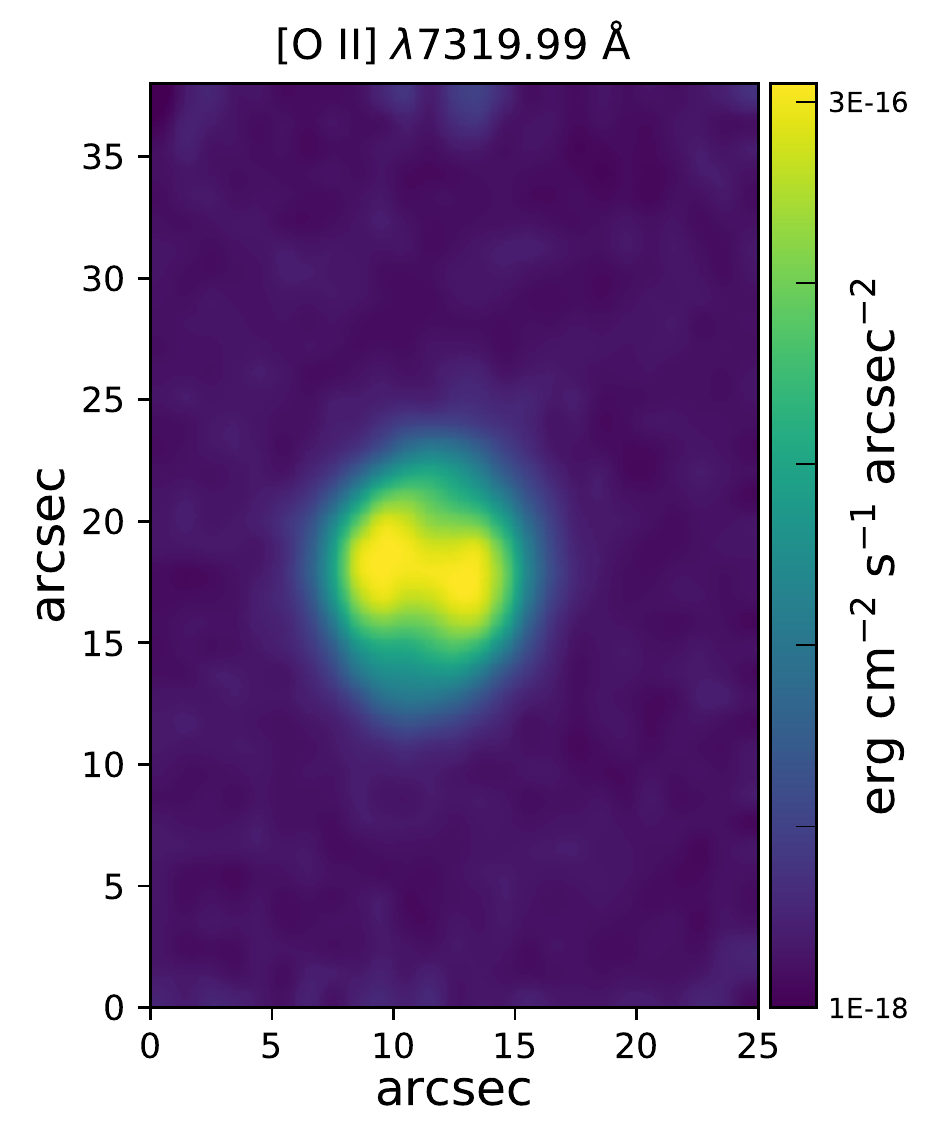} &
			\includegraphics[scale=0.60]{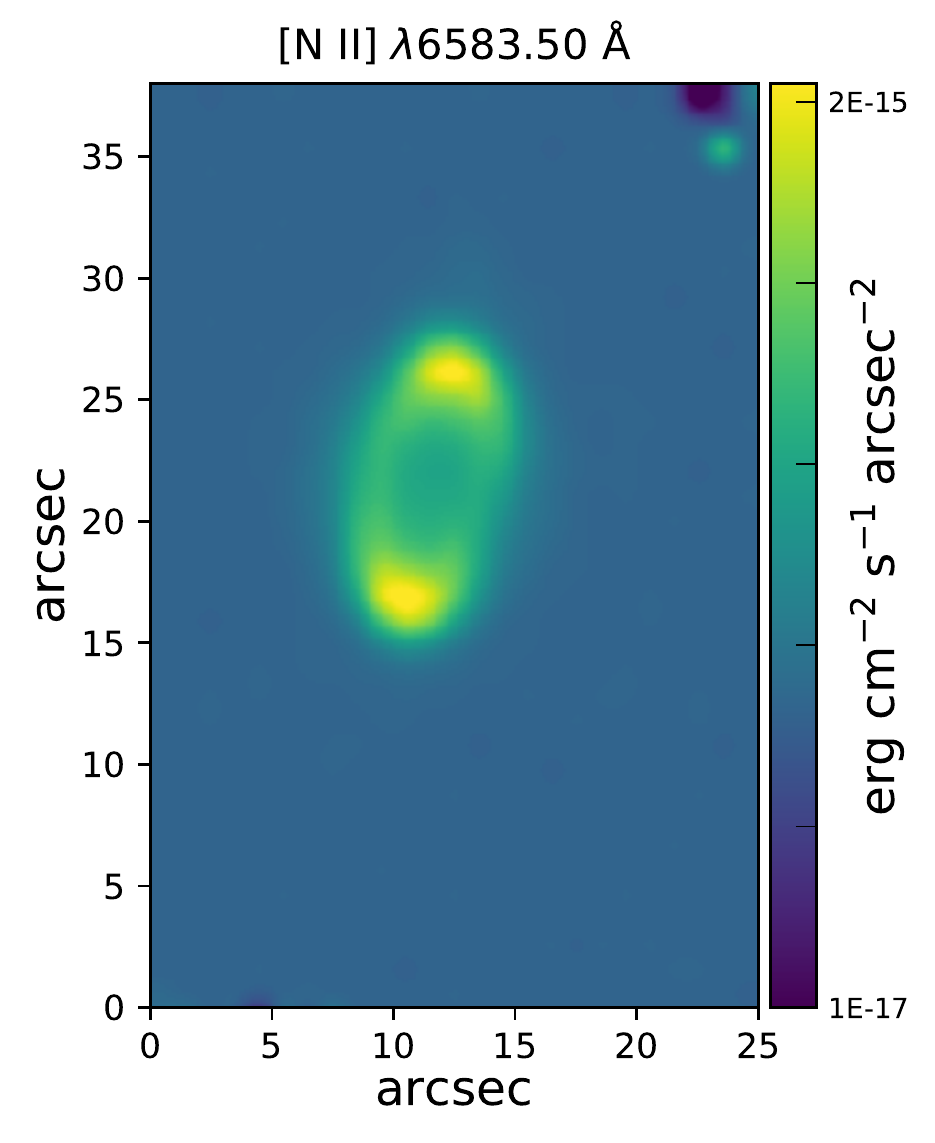}\\
            \includegraphics[scale=0.60]{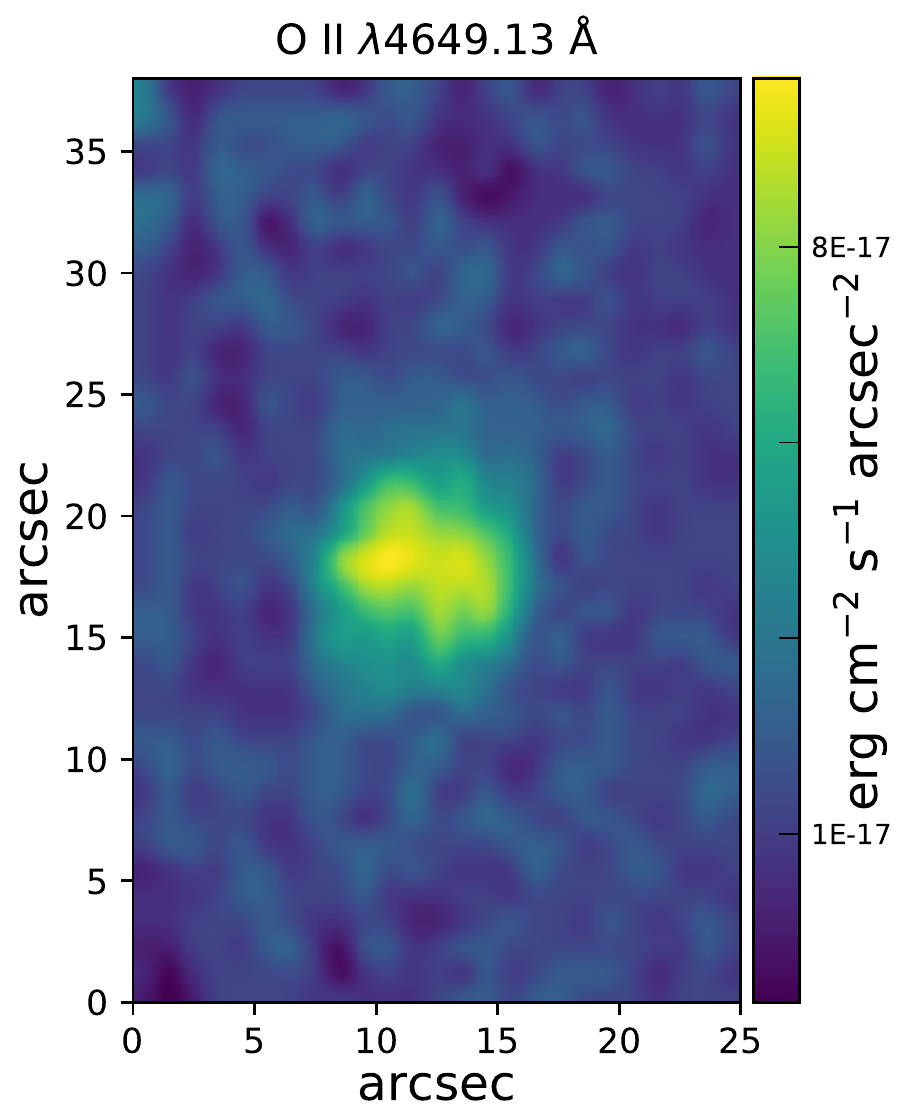} &
            \includegraphics[scale=0.60]{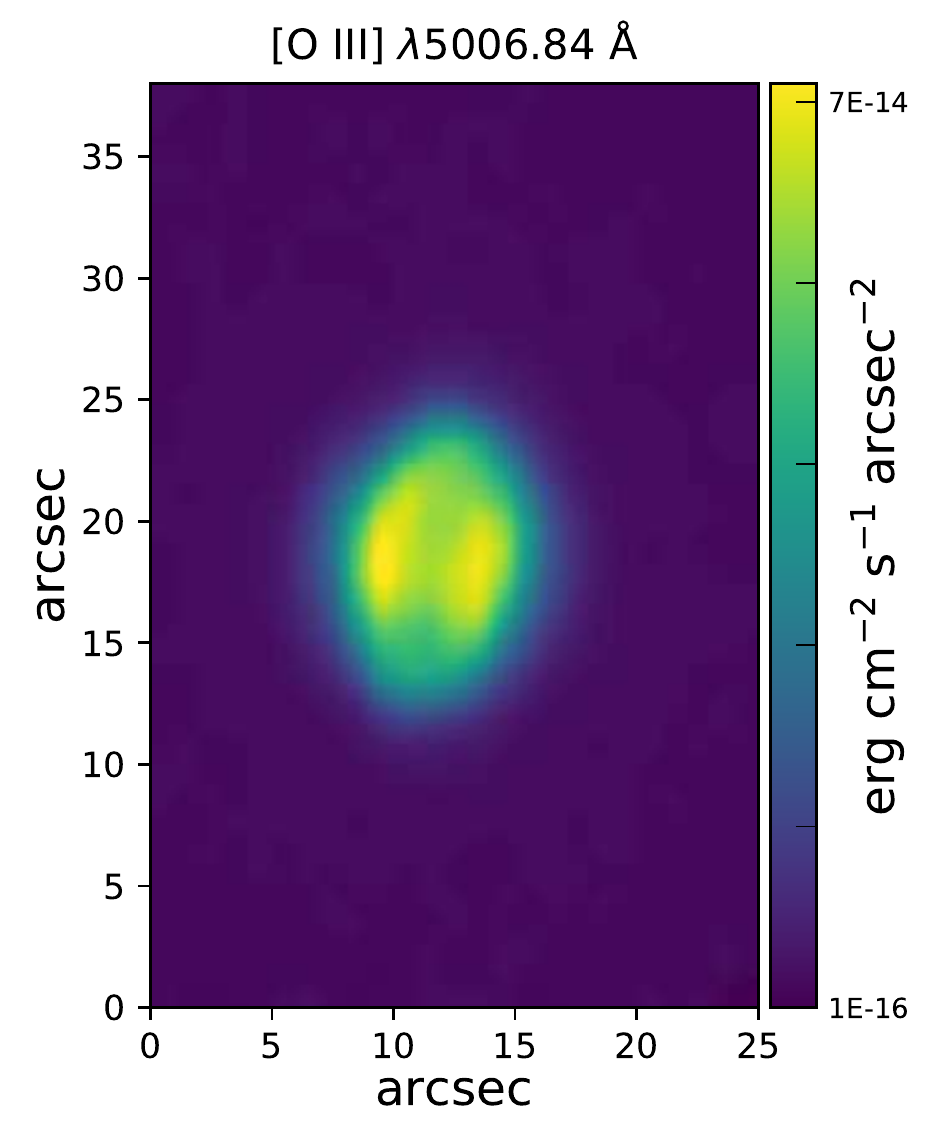} &
			\includegraphics[scale=0.60]{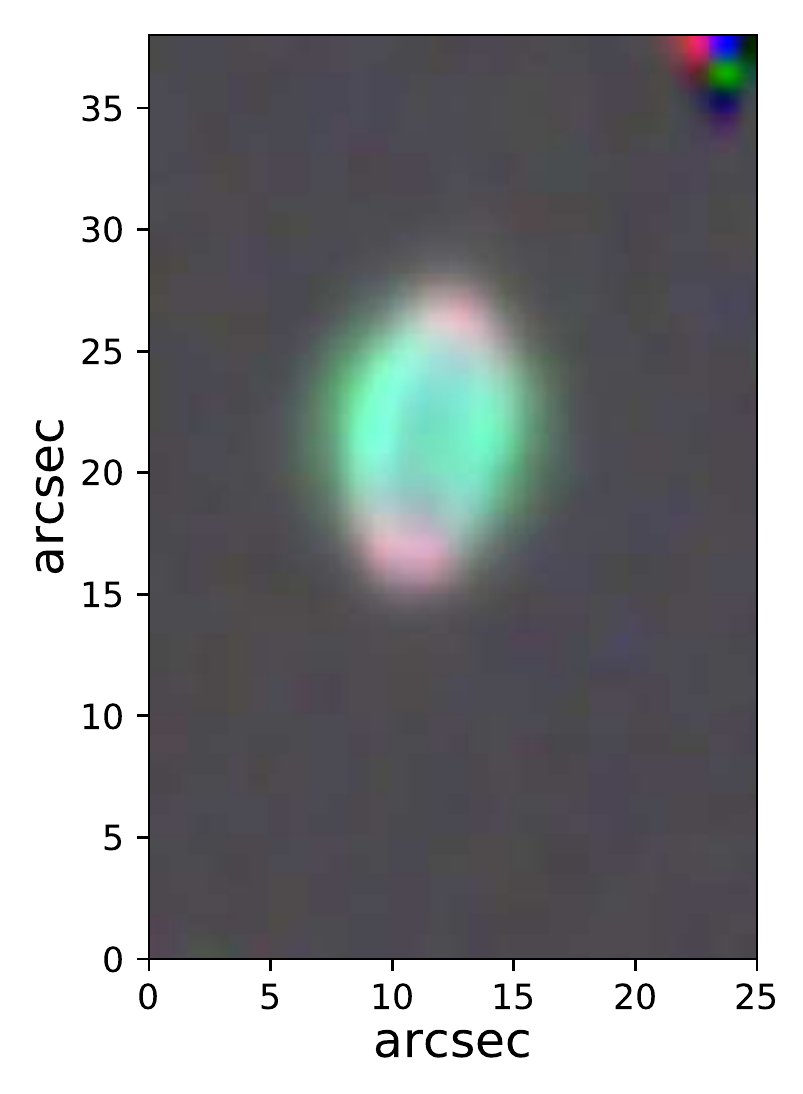}
	\end{tabular}}
	\caption{In upper panel, we present the PB\,4 emission-line maps of H$\gamma$ (left), [O\,II] (middle), and  [N\,II] (right).  In the lower panel, we present the maps of PB\,4 in O\,II (left) and [O\,III] (middle) emission lines. In right lower panel, we present an RGB composite colour image of PB\,4 in the lines of [SII]6730\,\AA (red channel), He I 6678\,\AA (green channel), and [S III] 6312\,\AA (blue channel). All images show an elliptical morphology for PB\,4, except the O\,II image. Further there are two opposite knots along the nebular major axis appear only in the [N\,II] image with yellow colour and RGB image with magenta (R+B) colour.} \label{Figure3}
\end{figure*}

To study the morphology and ionisation structure of the PNe sample, we extracted a number of emission-line maps representing different ionisation zones from their data cubes. In Figure \ref{Figure1}, we present three images of IC\,2501 in the [S\,III], [S\,II], and [Ne\,III] emission lines. The nebula displays a featureless roughly elliptical shape with a fainter outer shell visible  in the green and blue colours. The outer shell is only marginally distinguishable from the background sky emission.

Figure \ref{Figure2} displays three surprisingly different morphologies of the Hen 2-7 nebula in six emission-line maps. This shows a butterfly shape with an equatorial density enhancement forming a dense torus of gas around the CS, best seen in the [N\,II], [S\,II], and [O\,II] lines. The two lobes are oriented at $PA\sim 120$ with somewhat brighter emission in SE lobe compared to NW lobe. The object shows an elliptical morphology in the [O\,III] and He\,II emission lines with a roughly round central dense emission. In general, the surface brightness decreases in both maps from the centre to the outer boundary.

The [Fe\,III] emission-line map is remarkable, showing a jet-like morphology with point-symmetric condensations of emissions along the polar axis. Purely from the point of view of photoionisation theory, the morphology of [Fe\,III] should be similar to that of [O\,II], since the ionisation potential of [Fe\,II] is 16.19\,eV, that of [Fe\,III] is 30.65\,eV while the ionisation potential of [O\,I] is 13.62\,eV and that of [O\,II] is 35.12\,eV. The great difference of morphology between the  [Fe\,III]  and [O\,II] emission can only be explained by invoking a jet, either excited through shocks in a jet, or in a mass-loaded polar-directed stellar wind. The idea that the [Fe\,III] emission arises in a jet is confirmed by the kinematics, since relative to the systemic velocity, the SE jet is approaching at $+51\pm5$\,km/s at its tip, while the NW jet is receding at $-44\pm5$\,km/s. The strength of the [Fe\,III] emission in the jets strongly suggests that the Fe-containing dust has been destroyed within them. This points to the origin of the outflowing gas being close to the CS, within the dust sublimation radius, suggestive of a mass-loaded outflow from an interacting binary CS.

Figure \ref{Figure3} displays the morphology of PB\,4 nebula in five different emission-line maps and one composite RGB colour image. In general, the object shows an elliptical morphology, except O\,II map, with two symmetric maximum of brightness along the minor axis (likes a barrel with open ends) in the H$\gamma$, [O\,II], and [O\,III] maps. In the [N\,II] emission-line map, the nebula has an elliptical ring-like shape with two symmetric knots, in yellow colour, located at the ends of the nebular major axis. The [S\,II]-He\,I-[S\,III] composite colour image of PB\,4 shows the same morphology that is seen in the [N\,II], with the two symmetric knots which appear in magenta in the (B+R) colour image.

\subsection{Does PB\,4 host a close binary central star?}\label{binary}

\citet{Corradi15} have reported a possible relation between the largest adfs and the CS binarity of PNe. They studied three nebulae (A\,46, A\,63, and Ou\,5) with a post common-envelope binary star. A\,46, and Ou\,5 show $\rm O^{2+}/H^{+}$ adfs larger than 50, and this range as high as 300 in the inner regions of A\,46. A\,63 shows a smaller adf around 10. \citet{Jones15, Jones16} have strengthened the case for a  correlation between  elevated or extreme adfs and CS binarity in PNe through their study of Hen 2-155, Hen 2-161, and NGC 6778.  \citet{Sowicka17} suggested a possible correlation between low adfs and intermediate-period post-common-envelope CSs. Very recently \citet{Wesson18} confirmed the link between CS binarity and extreme adfs in PNe. They presented deep spectra for seven PNe host close binary CSs and found a several of them with extreme adfs. Further they analysed a small statistical sample of 15 PNe showing elevated or extreme adfs and found no link between CS surface chemistry and nebular discrepancy, but a clear link between binarity and the abundance discrepancy. Their analysis also revealed an anti-correlation between abundance discrepancy and nebular electron densities. Furthermore this study discovered that the PNe with binary CSs of a period $<$1.15 d have adfs exceeding 10, and an electron density less than $\sim$ 1000\,cm$^{-3}$ while the longer period binaries have adfs $<$10 and much higher densities. Thus they conclude that the adf can be used as a trusted tool in identifying the presence of a binary CS. We might speculate that this is because the UV spectrum of the companion star induces strong fluorescent pumping of the recombination lines, as found in the case of the relatively cool central star of IC\,418 by \citet{Morisset09}. Other possible explanations are discussed below.

The CS of PB\,4 nebula was classified as a true WELS type \citet{Weidmann15}, but we proved her this classification is in erroneous (see Section 5) due to most WELS feature lines arise in the nebula. \citet{Miszalski11} and references therein, pointed out that many of the characteristic WELS emission lines are observed in close binary CSs of PNe. Based on the morphology of a sample of 458 PNe, \citet{Soker97} introduced a classification scheme to characterize the evolution of binary progenitors of PNe. The basis of the classification is that axisymmetrical, bipolar, and elliptical PNe morphology arise from axisymmetrical mass loss from progenitor star as a result of the interaction of the progenitors with binary companion where the companion can be stellar or substellar (brown dwarfs or planets). According to this scheme, the progenitor star of PB\,4 was classified (with low confidence) as a progenitor which is interacting with substellar companion.

Our results show that PB\,4 nebula has a particularly strong recombination line spectrum (Table \ref{Table2}) with de-reddened line fluxes of 6.1, 3.2, and 2.2 for C\,II $\lambda$4267.15,  N\,III $\lambda$4640.64, and O\,II $\lambda$4649.13 on the scale of F(H$\beta$)=100, respectively. A comparison between the oxygen and nitrogen abundances computed from the ORLs and CELs, confirms that PB\,4 displays an extreme adf.

The map in O\,II line at $\lambda$4649.13 (Figure \ref{Figure3}) shows that the spatial distribution of $\rm O^{2+}$ recombination emission roughly matches those of $\rm C^{2+}$ $\lambda$4267.15 and $\rm N^{3+}$ $\lambda$4640.64, despite the much lower signal to noise in these other ions (Figure \ref{Figure4}). However, the O\,II $\lambda$4649.13 map has a spatial distribution which is remarkably different than that of the supposed parent ion emitting in the [O\,III] $\lambda$5006.84 CEL for which the emission is distributed through the entire object (Figure \ref{Figure3}). A similar result was achieved by \citet{Garcia16} where they found the O\,II ORL emission in NGC\,6678 is concentrated inside the [O\,III] CEL emission or H{~\small I} emission. Paradoxically, the O\,II $\lambda$4649.13 map much more closely resembles the [O\,II] $\lambda$7319.99 map, strongly suggesting that the O\,II ``recombination" lines do not arise from recombination at all, but from  fluorescent pumping of excited states in the  $\rm O^{+}$ ion. The fluorescent path for pumping the O\,II $\lambda$4649.13 transition is from the ground state through the UV3 $\lambda430$ line to the 3d$^4{\mathrm P}$ state, which can then decay via the V11, V19 and V28 multiplets. The path through the V11 multiplet then decays through the  O\,II $\lambda$4649.13 line.

In PB4, the fact that the O\,II $\lambda$4649.13 and the [O\,II]  $\lambda$7319.99 maps are extended in the polar direction, while the [N\,II] $\lambda$6583.5 map shows only an elliptical ring seems to show that the EUV radiation which is doing the pumping is preferentially directed in the polar direction, suggesting that the central binary is an interacting system, which would be consistent with a short-period binary CS, according with the conclusions of the \citet{Wesson18} study. The presence of a fairly hot binary companion is further suggested by the relatively strong continuum emission seen in our spectra. Apart from the NaD interstellar absorption lines, this spectrum is featureless. This enhanced stellar continuum results in the derivation of a very low Zanstra Temperature of 46000\,K \citep{Gleizes89}, while our photoionisation modelling (presented below) results in a stellar effective temperature $T_{\mathrm{eff}} = 93000\pm 3000$\,K.

In conclusion, we infer that the CS of PB\,4 is almost certainly an interacting binary with a sub-dwarf O-type companion, and that a poleward directed EUV continuum produced in the interaction is fluorescently pumping excited states in the  $\rm O^{+}$ ion to produce the extreme adf seen in the polar regions of this object. However, a photometric variability study and spectroscopic radial velocity observations would be essential to confirm the  binary nature of the central star.

\subsection{Expansion velocities}
The expansion velocities of the sample were measured, following \citet{Gieseking86}, from a number of emission lines of different ionisation levels. The ionisation potential (IP)  of each emission line is presented alongside the expansion velocity (V$_{exp}$) in Table  \ref{Table7}. It is obvious from the results that there is a general trend for increasing  expansion velocity with decreasing the ionisation potential, as would be expected from the ionisation stratification of these nebulae.

reviewing the literature, no measurements are found for IC\,2501 and PB\,4. Hen\,2-7 has three velocity measurements of 15.3, 18.3, and $<$6 km/s derived from [O\,III], [O\,II] and He\,II emission lines, respectively \citep{Meatheringham88}. The uncertainties in the previous values were not calculated by the authors but they reveal a statistical errors of $\sim$ 10\% in the measurements of the full width at half-maximum. Our velocity measurements of Hen\,2-7 (Table \ref{Table7}) indicate higher values compared with those given by \citet{Meatheringham88}. This could be attributed to the different spectral resolution and the different spectroscopic techniques which have been employed.

\subsection{Distances} \label{Distance}
None of IC\,2501, Hen\,2-7, and PB\,4 has a reliable individual distance based upon trigonometric parallax, spectroscopic parallax of the CS, cluster membership, or angular expansion distances.  Therefore, we must rely here on the statistical distance methods. We adopted the average distance deduced from the two recent statistical distance scales of \citet{Frew16} and \citet{Ali15a}. The results are given in Table \ref{Table3}. On the basis of these distances, we derived absolute luminosities $L_{\rm H\beta}$: $6.82\times10^{34}$~erg~s$^{-1}$, $1.33\times10^{34}$~erg~s$^{-1}$, and $1.22\times10^{34}$~erg~s$^{-1}$ for IC\,2501, Hen\,2-7, and PB\,4 respectively. However, the detailed photoionisation modelling is capable of providing an independent distance estimate, as discussed in \citet{Basurah16}. Based upon these, improved distance estimates are presented below.

\begin{table*}
	\centering \caption{Measured expansion velocities of IC\,2501, Hen\,2-7, and PB\,4.} \label{Table7}
	{
		\begin{tabular}{lcllll}			
			\hline
			Ion		&	$\lambda$ (\AA)			&	IP		& \multicolumn{3}{c}{V$_{exp}$ (km/s)} \\
			\cline{4-6} \\
			&				&			&	IC\,2501	&	 Hen\,2-7	&	PB\,4	\\
			\hline												
			{[Ne\,III]}	&	3868 ~\& 3967	&	63.45	&	6.86	&	21.50	&	17.90	\\
			{[Ar\,IV]}	&	4740	&	59.81	&	9.20	&	21.30	&	17.30	\\
			{[O\,III]}	&	4959  ~\& 5007	 &	54.93	&	11.40	&	22.00	&	18.64	\\
			{[Cl\,IV]}	&	7530	&	53.46	&	12.30	&	22.10	&	17.11	\\
			{[Ar\,III]}	&	5191 (7751)	&	40.74	&	12.27	&	24.98	&	18.90	\\
			{[Cl\,III]}	&	5517 ~\& 5537	&	39.61	&	15.59	&	26.90	&	19.62	\\
			{[O\,II]}	&	3727 ~\& 3729	&	35.12	&	17.91	&	31.40	&	19.36	\\
			{[N\,II]}	&	6548 ~\& 6583	&	29.6	&	18.64	&	31.09	&	18.35	\\
			{[S\,II]}	&	6716 ~\& 6731	&	23.33	&	18.89	&	30.79	&	18.70	\\
			{[N\,I]}	&	5197 ~\& 5200	&	14.53	&	19.74	&	34.00	&		\\
			{[O\,I]}	&	6300 ~\& 6363	&	13.62	&	23.20	&	33.10	&		\\
			\hline											
			Average &                   &           &    15.10  & 27.20     & 18.43 \\
			\hline	
			
	\end{tabular}}
\end{table*}

\begin{figure*}
	{ \begin{tabular}{@{}ccc@{}}
			\includegraphics[scale=0.60]{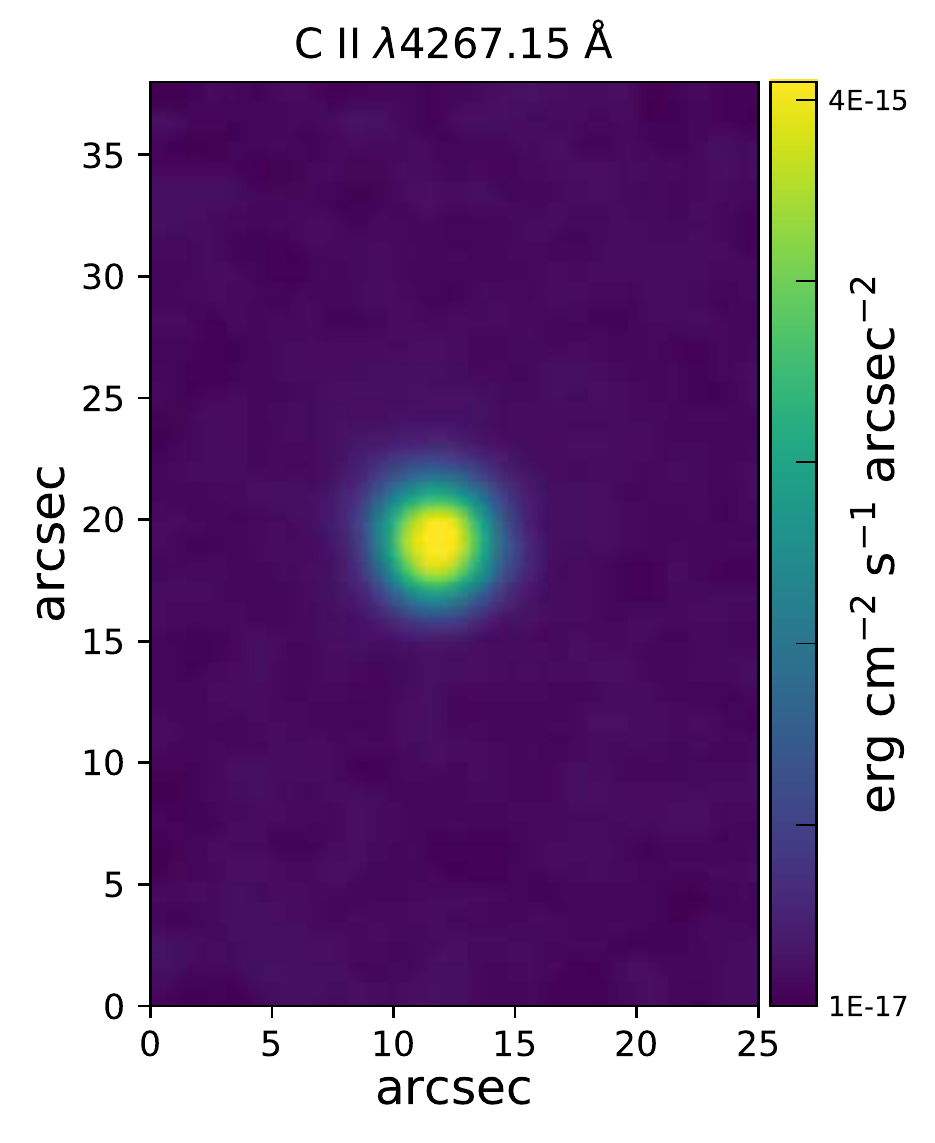} &
			\includegraphics[scale=0.60]{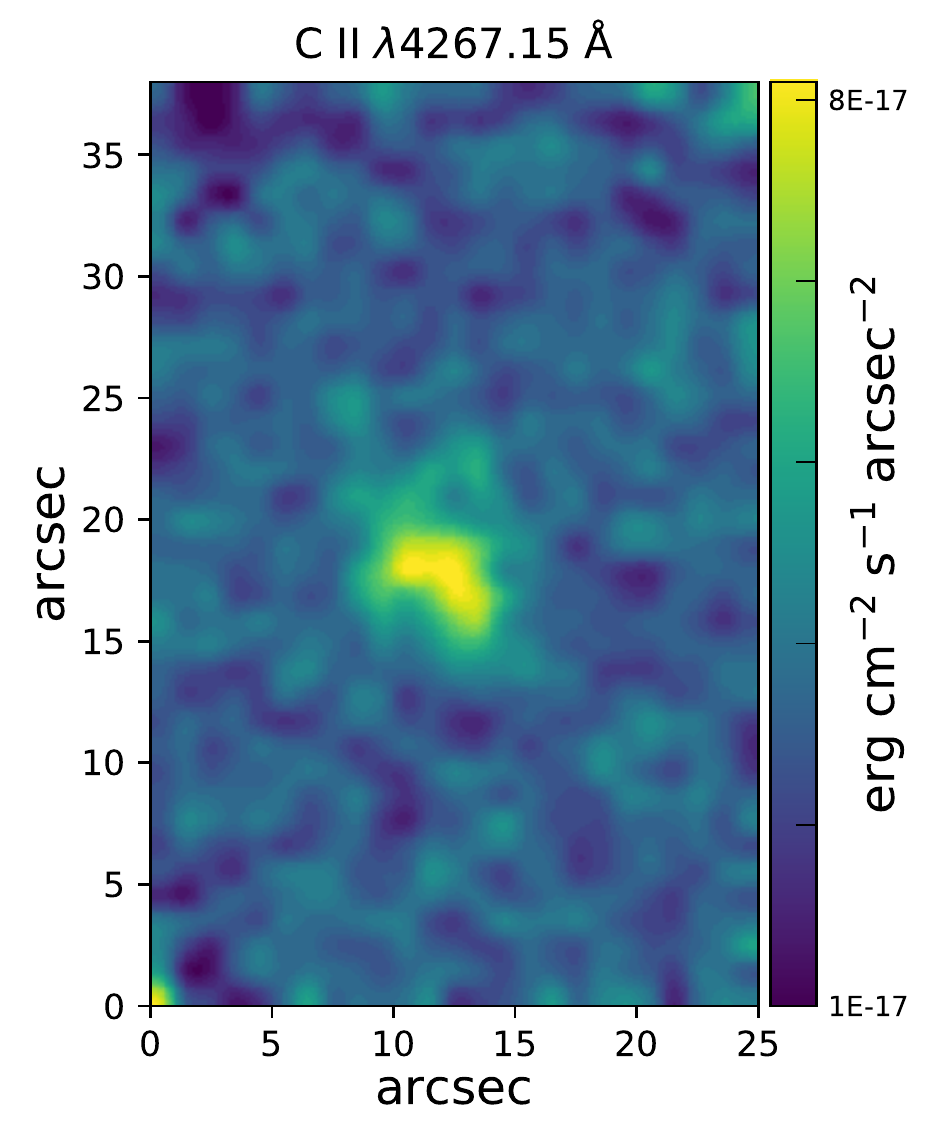} &
			\includegraphics[scale=0.60]{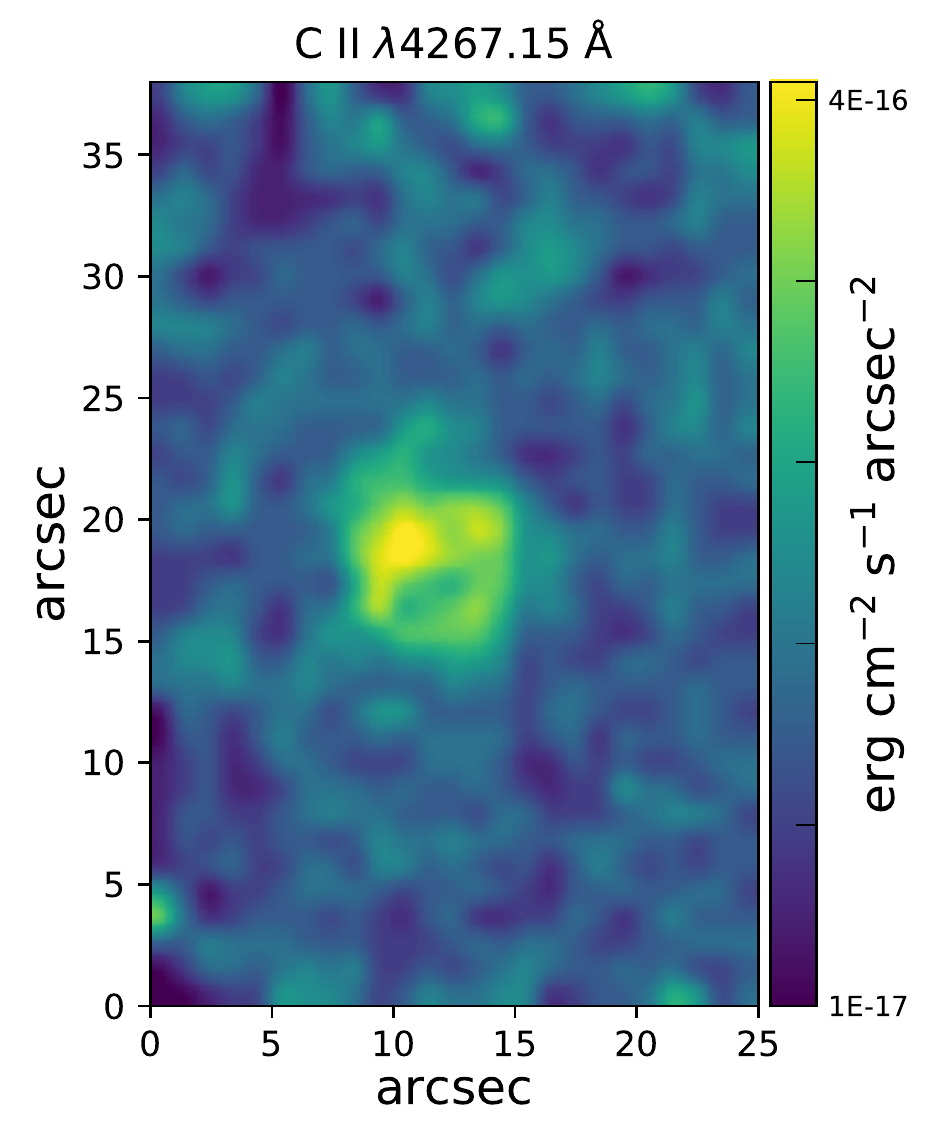} \\
			\includegraphics[scale=0.60]{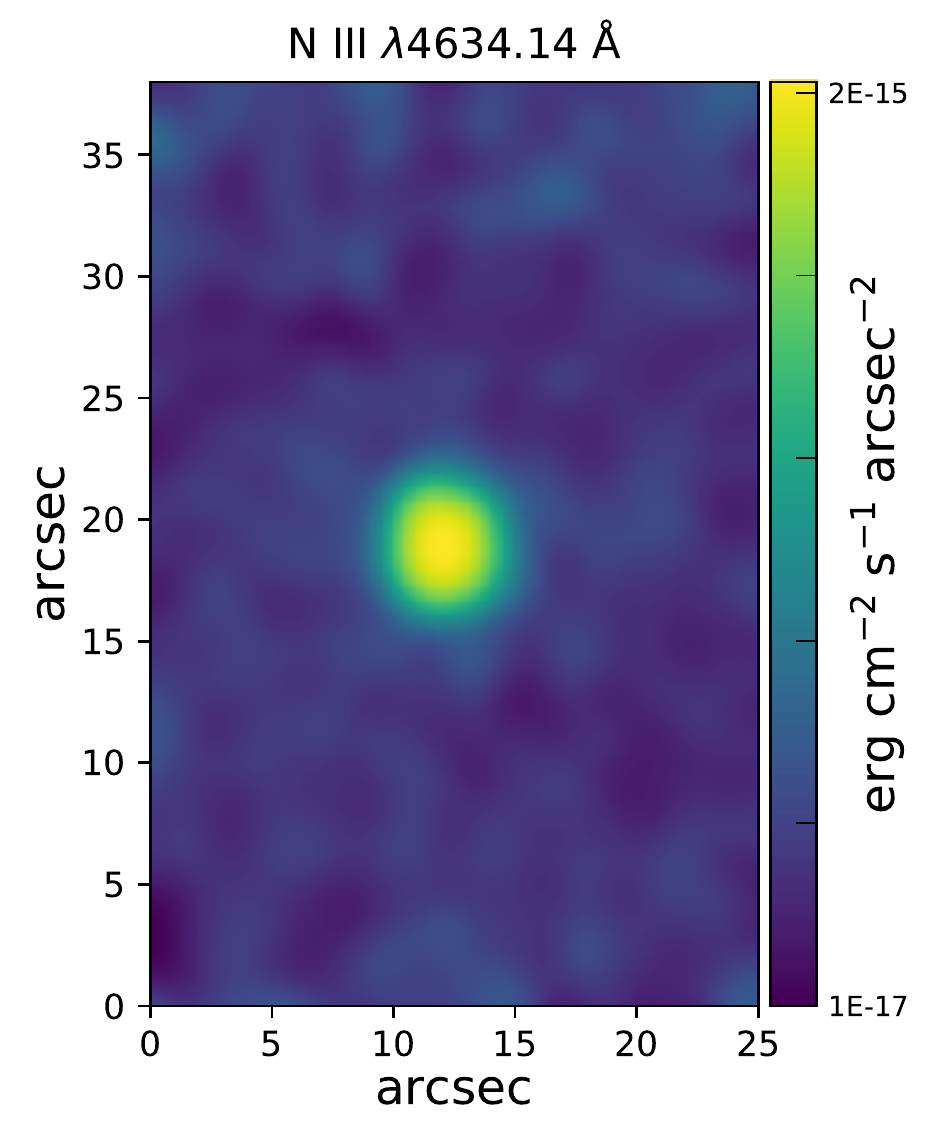} &
			\includegraphics[scale=0.60]{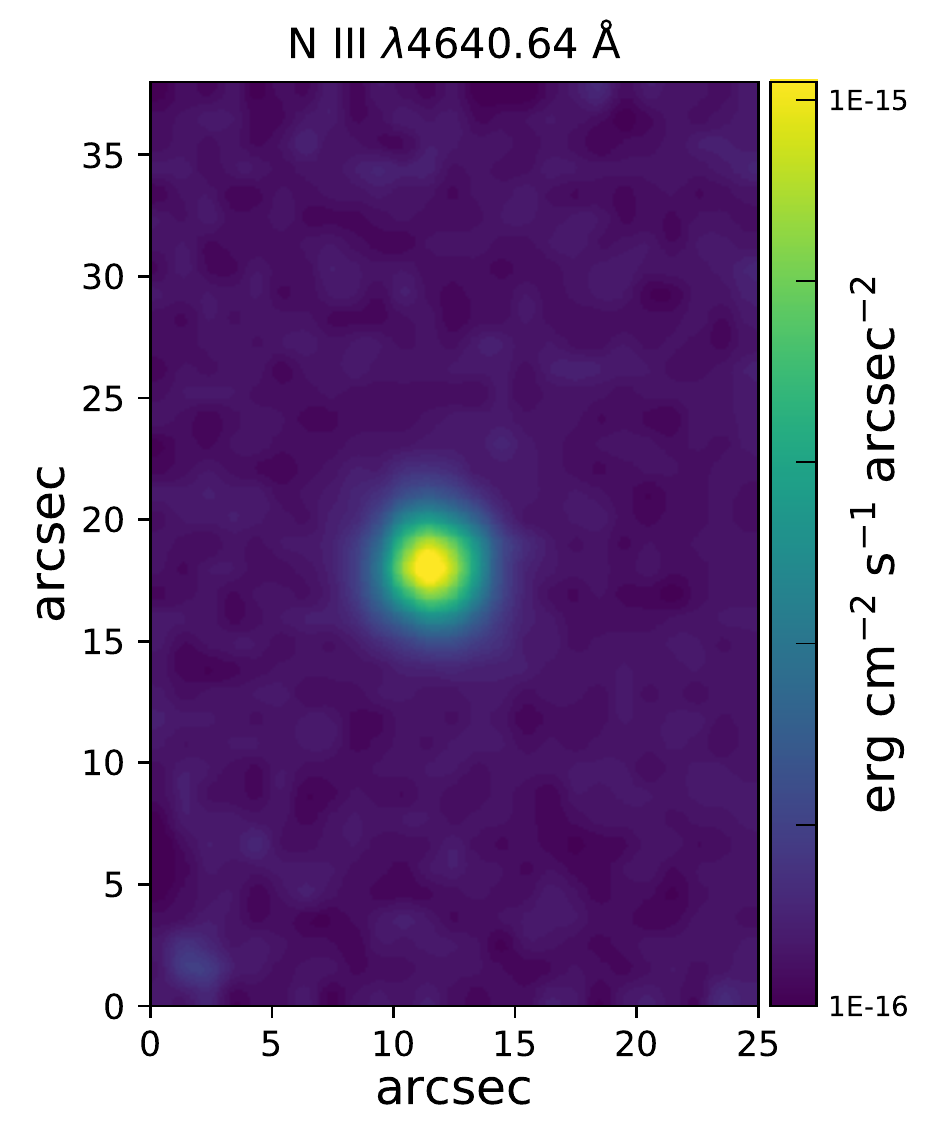} &
			\includegraphics[scale=0.60]{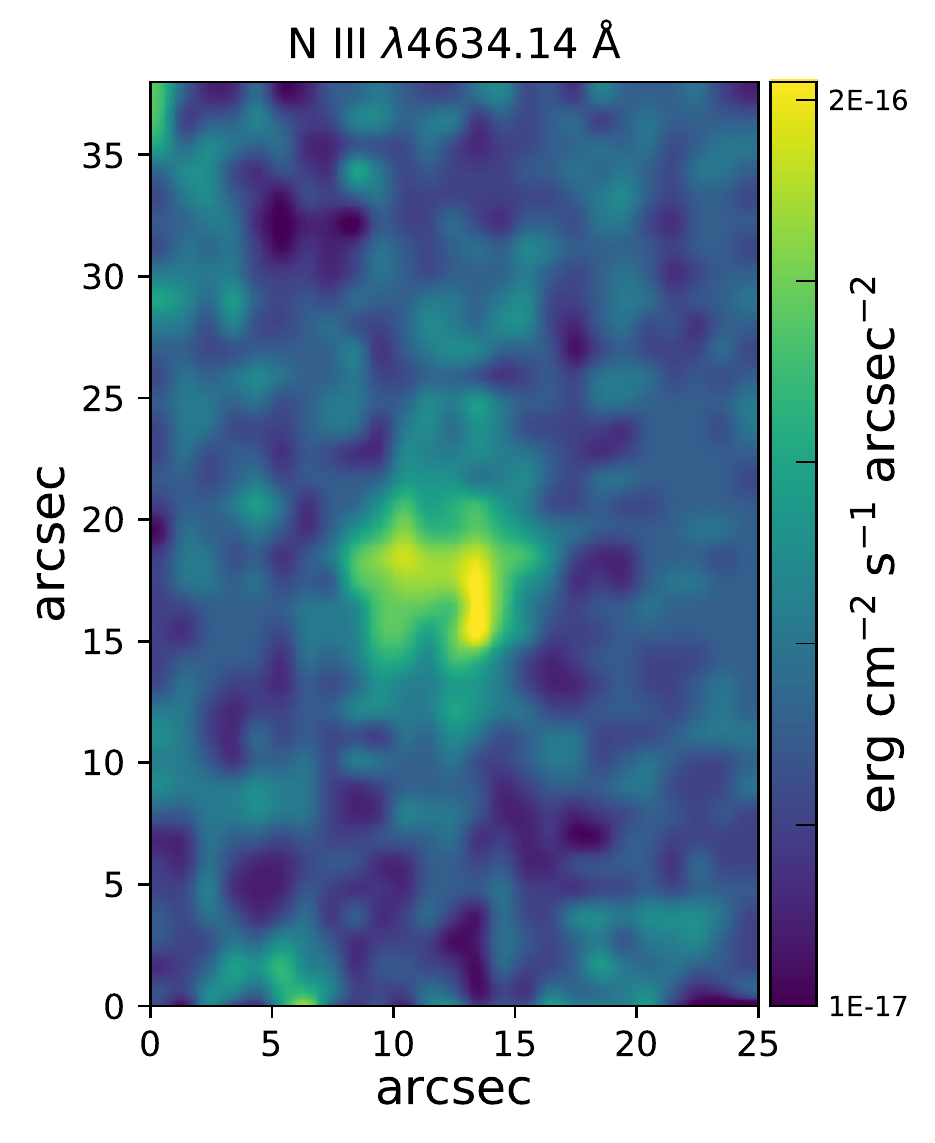}
	\end{tabular}}
	\caption{This figure shows the emission-line maps of IC\,2501 (left upper and lower panels), Hen\,2-7 (middle upper and lower panels), and PB\,4 (right upper and lower panels) in the optical recombination lines C\,II (upper panels), and N\,III (lower panels). Both lines are supposed to be of central star origin according to the WELS classification. However, it is obvious the emissions of both lines are spatially extended in all objects, and therefore of nebular rather than CS origin.} \label{Figure4}
\end{figure*}

\section{Misclassification of the IC\,2501 and PB\,4 central stars}

Through the analysis of $\sim 80$ emission-line CSs, \citet{Tylenda93} found that about half  belong to the  [WR] star class while others display emission lines at the same wavelengths as [WR] stars but having narrower line widths and weaker line intensities. They termed the latter group Weak Emission Line Stars (WELS). The spectral lines which characterise the WELS class are the recombination lines of C and N. These lines are  N\,III $\lambda \lambda 4634, 4641$,  C\,III $\lambda 4650$, C\,IV $\lambda 4658$, and C\,IV $\lambda \lambda  5801, 5812$. The emission line of C\,III $\lambda$5696 is either very weak or absent.

It is hard to discriminate the width of these lines from the nebular lines, although on low dispersion spectra the group of lines around 4650\AA\ and the C\,IV doublet around 5805\AA\ may appear somewhat broader. \citet{Miszalski09} claimed that many of WELS are likely to be misclassified close binaries. Further, \citet{Miszalski11} indicated that many of the characteristic WELS emission lines have been observed in close binary central stars of PNe and originate from the irradiated zone on the side of the companion facing the primary. \citet{Basurah16} discovered that, for few objects, the WELS classification may well be specious, since the WELS features originated in the nebula, rather than in the central star.  Additionally, in \citet{Ali16}, we successfully extracted the CS spectrum of the M3-6 nebula from its 3-D data cube. On the basis of the observed CS spectral lines, we re-classified it as a hydrogen rich star of spectral type O3 I(f*) rather than its prior WELS type.

The appearance of the proposed WELS features in the low dispersion (3.4\,\AA $/$pixel) CSs spectra of IC\,2501 and PB\,4  encouraged \citet{Weidmann11} and \citet{Weidmann15} to classify both CSs as probable WELS and true WELS types, respectively. Very recently, \citet{Weidmann18} re-classified the CS of IC\,2501 as O-type star. Further, they suggested that the CS could be a binary system due to the presence of a shift between the nebular emission lines and their corresponding stellar features. This result supports the above conclusion that most weak emission-line stars are indeed close binary stars. Inspecting the literature, the CS of Hen\,2-7 was not detected yet.

Unfortunately we are unable to extract the CS spectra at adequate S/N for any of this PNe sample from their data cubes, but it is clear that the WELS classification for IC\,2501 and PB\,4 are erroneous for these nebulae. In Figure \ref{Figure4}, we present emission-line maps of IC\,2501, Hen\,2-7, and PB\,4 in two proposed CS recombination lines of WELS type. It is obvious that the emission of both lines are spatially distributed over a large nebular area, therefore they are of nebular rather than CS origin.

\section{Self-consistent photoionisation modelling}
The details of our self-consistent photoionisation modelling have been adequately presented in our earlier papers \citep{Basurah16,Dopita17}, so we will only present a brief summary here. For the photoionisation modelling we use  {\tt Mappings V version 5.1.12} code \citep{Sutherland17}. {\footnote{Available at {\url{https://miocene.anu.edu.au/mappings}}}  For the UV spectrum of the central star, we use the  model atmospheres from \citet{Rauch03}. We assume that heavy elements are depleted onto dust. For the initial depletion factors we use the \citet{Jenkins09} scheme, with a base depletion of Fe of  -1.00\,dex, but these are adjusted by individual element to best fit the observations.  We allow for both grain charging, and photoelectron emission, which can be an important heating process in these nebulae  \citep{Dopita00}.

Our object is to use the de-reddened integrated global spectrum of the PNe to build a model which optimises the fit to a many of the observed emission lines as possible, while being consistent with the morphological properties of the nebula. This optimisation is done by minimising the L1-norm for the fit. That is to say that we measure the modulus of the mean logarithmic difference in flux (relative to H$\beta$) between the model and the observations \emph{viz.};
\begin{equation}
{\rm L1} =\frac{1}{m}{\displaystyle\sum_{n=1}^{m}} \left | \log \left[ \frac{F_n({\rm model})} {F_n({\rm obs.)}} \right]  \right |. \label{L1}
\end{equation}
This procedure weights fainter lines equally with stronger lines, and is therefore more sensitive to the values of the input parameters.

At the same time, we seek to match both the observed physical size and the absolute H$\beta$ flux, assisted by the physical appearance and structure of the nebula in the IFU emission line images. In general, the models consist of an optically thick component (in which lines such as [O\,I], [O\,II], [N\,I], [N\,II], and [S\,II] mostly arise), as well as an optically-thin component which gives rise to emission lines of He\,II [O\,III],  [Ne\,III][, [Ar\,III] as well as lines of higher ionisation stages. The schematic geometry of our model is shown in Figure \ref{model}. The outer radius of the optically thick zone is determined by the ionisation parameter at the inner boundary $U_{in}$, the EUV luminosity of the central star and the nebular pressure $P$. For the optically thin region, the outer radius is determined by these three parameters as well as the adopted optical depth at the Lyman limit, $\tau$.

\begin{figure}
\centering
	\includegraphics[scale=0.3]{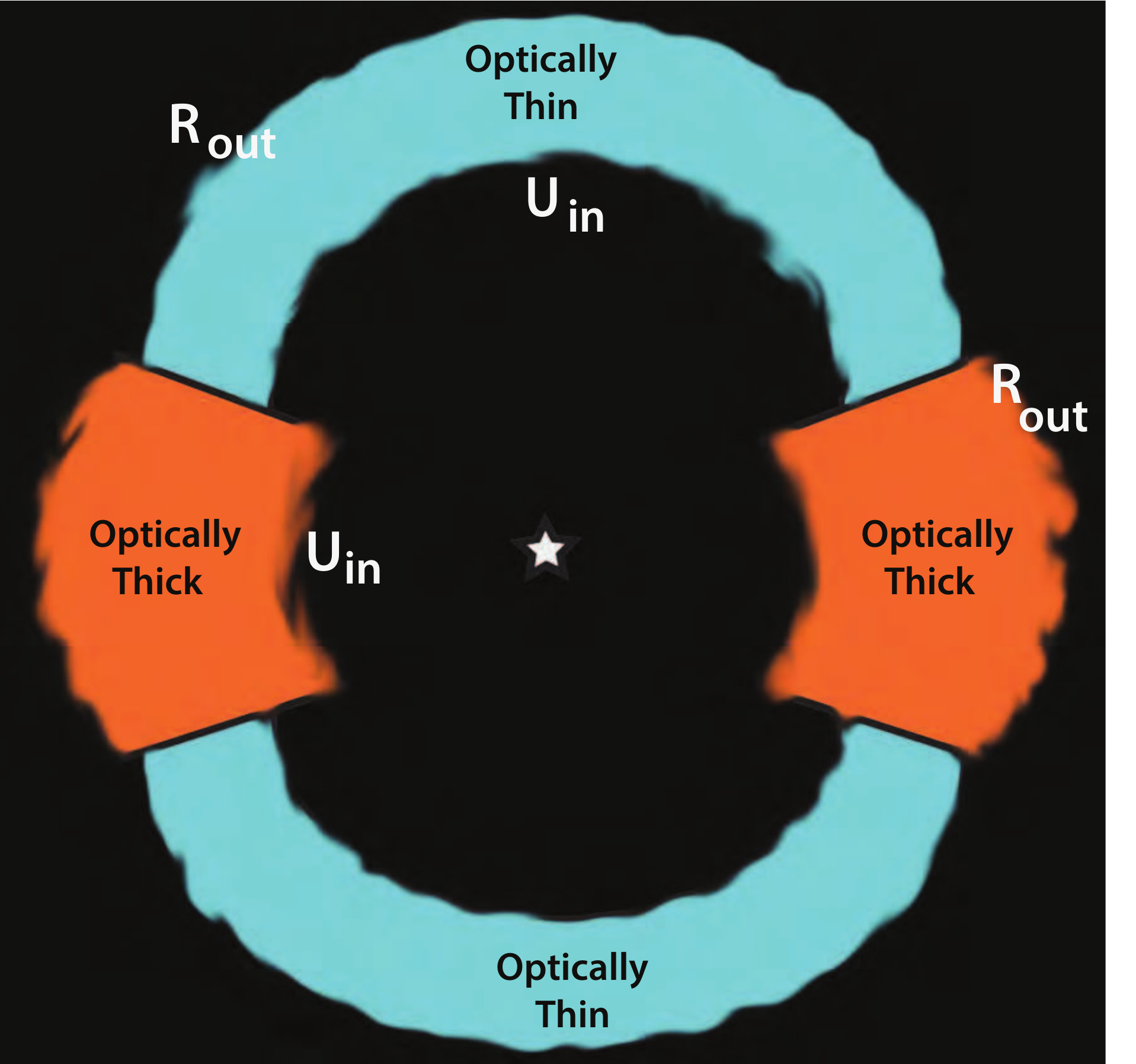}
	\caption{The geometry of our detailed photo-ionisation model. The diagram represents a slice through the mid-plane of the nebula, which has rotational symmetry. The optically-thick portion is therefore a torus in the equatorial plane.} \label{model}
\end{figure}

The fractional contributions of the optically-thick and optically-thin components are determined by line ratios such as He\,II$\lambda4686$/He\,I$\lambda5876$, and [O\,III]$\lambda5007$/H$\beta$, which are commonly use to define the Excitation Class (EC) of planetary nebulae. These ratios as also sensitive to the Effective Temperature ($T_{\mathrm{eff}}$) of the central star, but the ambiguity can be addressed by examination of the strengths (relative to H$\beta$) of the low-excitation species such as  [O\,I] , [N\,I] or  [S\,II], since in high excitation, high $T_{\mathrm{eff}}$ and optically thick nebulae, these lines would be relatively strong, while in optically thin nebulae with similar EC but with lower $T_{\mathrm{eff}}$ these lines are much weaker.

The pressure in the ionised plasma is determined by matching the density determined by the  density-sensitive [O\,II] and the [S\,II] lines in the optically thick component, and the [Cl\,III] and [Ar\,IV] lines in the high excitation zone; see Table \ref{Table4}. The electron temperature produced by the model should also match the values measured in Table \ref{Table4}, however, the temperature is strongly dependent on the chemical abundances adopted, and the very important carbon line cooling in the UV.

In the iterative L1-norm minimisation process, after we have have matched the nebular size, excitation, pressure, $T_{\mathrm{eff}}$ of the central star, and the H$\beta$ flux (which constrains the luminosity of the central star, $L_*$), we adjust the abundances of individual elements in order to provide an optimum fit for all observed ionisation stages of each element.

We now describe the detailed photoionisation fitting to each nebula.

\subsection{IC\,2501}
The images of IC\,2501 in Figure \ref{Figure1} show a smooth featureless elliptical shell. From the strength of the [O\,I] doublet $\lambda\lambda6300,6363$ it is evident that this nebula is optically thick. We have therefore modelled this nebula as a single optically-thick shell with a mean radius equal to the average of the major and minor axes. At the assumed distance of 2.64\,kpc, this corresponds to a radius of $1.55\times10^{17}$\,cm.

Given that the nebula is optically-thick, the extreme weakness of he He\,II $\lambda4686$ line provides a tight constraint on the stellar effective temperature, $50< T_{\mathrm{eff}} <60$\,kK. The detailed model fit gives $T_{\mathrm{eff}} =55\pm3$\,kK with an inner ionisation parameter $\log U_{\mathrm{in}} = -2.15$, corresponding to a nebular inner radius of $1.0\times10^{17}$\,cm. The stellar luminosity is $L_* = (1.15\pm0.2)\times 10^{37}$\,erg/s, or $2990\pm 520$ \,$L_{\odot}$.

The full photoionisation model parameters for this and the other nebulae are given in Table \ref{Table8}, and the derived chemical abundance sets are listed in Table \ref{Table5}. The abundances given here are the gas phase abundances, and do not include the faction of the heavy elements trapped in grains. To estimate the total (gas + solid phase) abundances, one should multiply the abundances listed in Table \ref{Table5} by the following factors: He 1.00, C 1.32, N 1.00, O 1.05, Ne 1.00, Ar 1.00 and Cl 1.00. Table \ref{Table4} gives the modelled temperatures and densities for the various ions observed.

\subsection{He2-7}
This nebula presents more of a challenge to model, thanks to its ``butterfly'' structure, and the evidence presented in Section \ref{maps} that it is powered by a strong bipolar outflow. In this object we treat the equatorial ring as one region, and the bipolar structure as a second region. These two regions have quite different densities (and hence gas pressures). This is clearly shown by a map of the density-sensitive ratio; [S\,II]$\lambda\lambda6731/6717$, presented in Figure \ref{He2-7-dens}. This map was prepared by extracting the individual line images using {\tt QTFitsView} as .fits files, using the {\tt{iraf}} {\tt{imrepl}} task to trim the [S\,II]$\lambda6731$ image to remove noisy pixels, and then dividing one image by the other using the {\tt{imarith}} task.

\begin{figure}
\centering
	\includegraphics[scale=0.50]{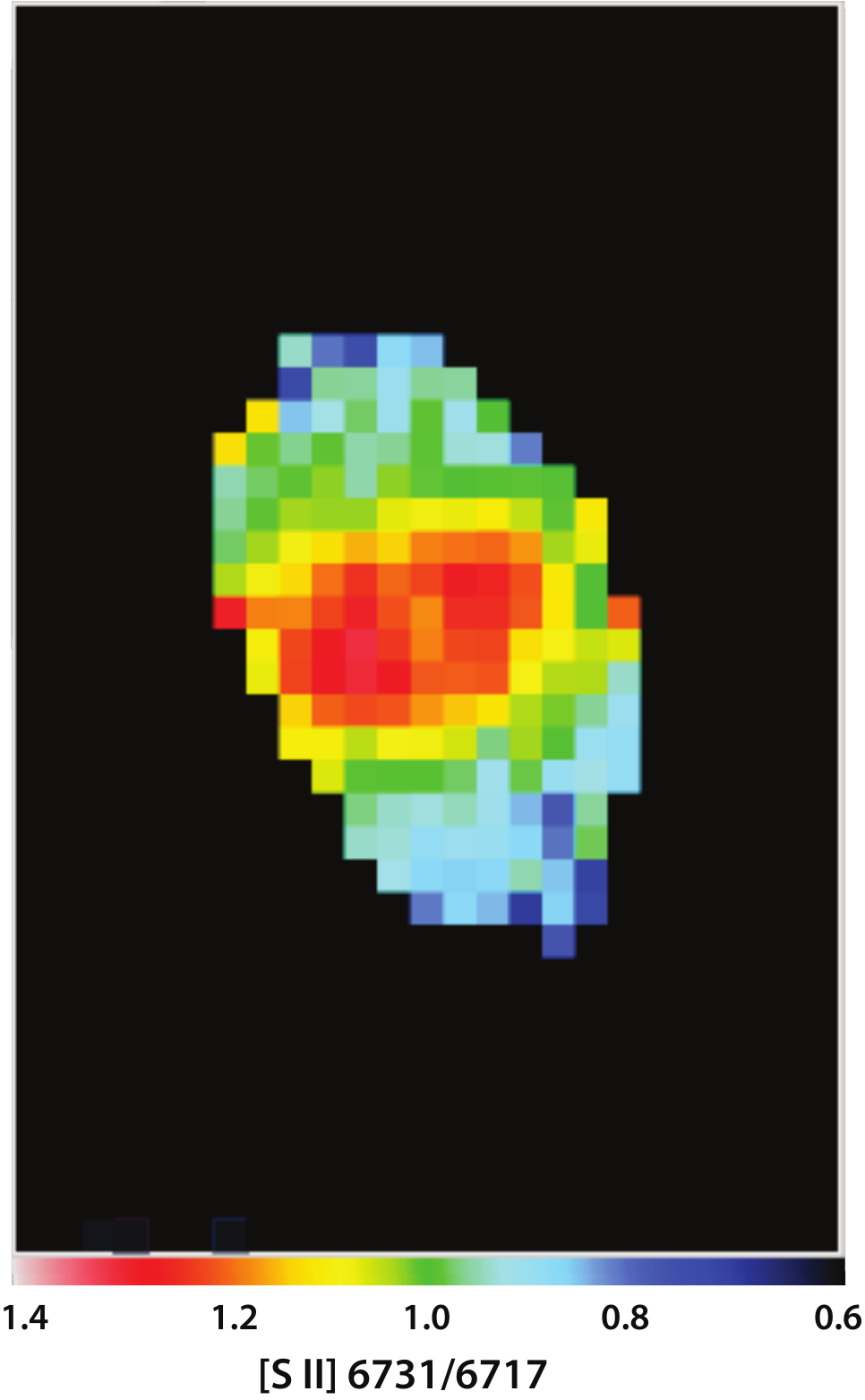}
	\caption{The spatial variation of the density-sensitive [S\,II] $\lambda\lambda6731/6717$ ratio in He\,2-7. It is evident that the elliptical ring is systematically denser than the ``butterfly'' wings of this bipolar nebula.} \label{He2-7-dens}
\end{figure}

From Figure  \ref{He2-7-dens} we deduce that the nebula can be modelled as an inner ellipse of radius 5.5 arc sec., and an electron density of $1050\pm130$\,cm$^{-3}$, while the outer butterfly wings at $PA\sim 120^o$ have a radius of 9.5 arc sec. and a mean electron density of $400\pm100$\,cm$^{-3}$. Again the He II line strength provides a good constraint on the stellar effective temperature; $T_{\mathrm{eff}} =83.5\pm1.5$\,kK.

In order to determine the luminosity, we used the procedure described in  \citet{Basurah16} by varying the assumed distance  to obtain a simultaneous fit for the nebular radius in the two zones of the model, as well as the absolute H$\beta$ flux. This procedure yields a somewhat larger distance than given in Table \ref{Table3}; $4.3\pm0.5$\,kpc, giving an outer radius of $3.3\times10^{17}$\,cm for the inner ellipse (\emph{vs.} $3.8\times10^{17}$\,cm for the model), and $6.3\times10^{17}$\,cm for the butterfly wings, \emph{vs.} $6.8\times10^{17}$\,cm for the model. With relative flux weighting factors of 0.32 for the central ellipse \emph{vs.} and 0.68 for the butterfly wings (which minimises the L1-norm of the fit), the H$\beta$ flux is predicted to be $\log L_{H\beta} = 34.35$, vs the observed value (at an assumed distance of 4.3\,kpc) of $\log L_{H\beta} = 34.3$. The implied luminosity of the central star is $L_* = (4.5\pm1.1)\times10^{36}$\,erg/s, or $1175\pm300 L_{\odot}$.

\begin{table}
	\centering \caption{Photoionisation model parameters for IC\,2501, Hen\,2-7, and PB\,4.} \label{Table8}
	  \scalebox{0.8}{
		\begin{tabular}{lll}			
			\hline
			\hline
			Nebula		&	Parameter	&	Value \\
			\hline
			{\bf IC\,2501}	& Stellar Luminosity,  $L_*$                               & $2990\pm 520$ \,$L_{\odot}$ \\
			                            & Stellar Effective Temperature, $T_{\mathrm{eff}}$ & $55000\pm3000$\,K \\
			                            & Estimated Distance                                           & $2.64\pm0.4$\,kpc \\
			                            & L1-norm of fit (dex.)		                         & 0.079 \\
			                            &                                                                              &  \\
			                            & \emph{Optically Thick Component:}              &   \\
			                            & Inner ionisation parameter, $\log U_{in}$    &  -2.15     \\
			                            & Nebular Pressure, $P/k$\,(cm$^{-3}$K )       & $1.85\times10^8$ \\
			                            & Outer Radius @50\% HII (cm)                        & $1.52\times10^{17}$ \\
			\hline	
			{\bf He\,2-7}	& Stellar Luminosity,  $L_*$                                & $1175\pm 300$ \,$L_{\odot}$ \\
			                            & Stellar Effective Temperature, $T_{\mathrm{eff}}$ & $83500\pm1500$\,K \\
			                            & Estimated Distance                                           &  $4.3\pm0.5$\,kpc \\
			                            & L1-norm of fit (dex.)		                         & 0.045 \\
			                            &                                                                               &  \\
			                            & \emph{Optically Thick Component:}              &   \\
			                            & Flux weighting factor                                        &  0.32 \\
			                            & Inner ionisation parameter, $\log U_{in}$     &  -1.0    \\
			                            & Nebular Pressure $P/k$\,(cm$^{-3}$K)         & $3.0\times10^7$ \\
			                            & Outer Radius @50\% HII (cm)                         & $3.86\times10^{17}$ \\
			                            &                                                                               &  \\
			                            & \emph{Optically Thin Component:}                &   \\
			                            & Inner ionisation parameter, $\log U_{in}$     &  -2.0    \\
			                            & Nebular Pressure $P/k$\,(cm$^{-3}$K)         & $9.0\times10^6$ \\
			                            & Outer Radius (cm)                                             & $6.90\times10^{17}$ \\
			\hline	
			{\bf PB\,4}	& Stellar Luminosity,  $L_*$                                         & $1300\pm 390$ \,$L_{\odot}$ \\
			                            & Stellar Effective Temperature, $T_{\mathrm{eff}}$ & $93000\pm 3000$\,K \\
			                            & Estimated Distance                                           &  $3.1\pm0.3$\,kpc \\
			                             & L1-norm of fit (dex.)		                         & 0.090 \\
			                             &                                                                               &  \\
			                            & \emph{Disc Component:}              &   \\
			                            & Flux weighting factor                                        &  0.57 \\
			                            & Inner ionisation parameter, $\log U_{in}$    &  -2.0    \\
			                            & Nebular Pressure $P/k$\,(cm$^{-3}$K)        & $4.5\times10^7$ \\
			                            & Outer Radius @50\% HII (cm)                       & $2.4\times10^{17}$ \\
			                            &                                                                             &  \\
			                            & \emph{Polar Component:}                &   \\
			                            & Inner ionisation parameter, $\log U_{in}$    &  -1.0    \\
			                            & Nebular Pressure $P/k$\,(cm$^{-3}$K)        & $2.8\times10^7$ \\
			                            & Outer Radius (cm)                                            & $1.9\times10^{17}$ \\
			\hline	
			
	\end{tabular}}
\end{table}

\begin{table}
	\centering \caption{The photoionisation model fits for IC\,2501, Hen\,2-7, and PB\,4.} \label{Table9}
	  \scalebox{0.8}{
		\begin{tabular}{lcrrrrrr}			
 \hline
 \hline
 &  &  \multicolumn{2}{c}{\bf{IC\,2501}}   &  \multicolumn{2}{c}{\bf{Hen 2-7}}  &  \multicolumn{2}{c}{\bf{PB\,4}} \\
 \hline
$\lambda$ & Ion & Flux & Model & Flux & Model & Flux & Model \\
\hline
\\
3727,9 & [O\,II] & 67.0 & 91.25 & 93.10 & 142.0 & 14.91 & 47.13 \\
3835 & H\,I & 6.54 & 7.34 & 7.12 & 7.33 & 6.62 & 7.34 \\
3869 & [Ne\,III] & 76.0 & 74.02 & 105.0 & 96.60 & 77.70 & 75.05 \\
3888 & He\,I & 18.20 & 13.40 & 18.80 & 12.12 & 19.80 & 14.48 \\
3967 & [Ne\,III] & 22.20 & 22.30 & 24.90 & 29.10 & 11.30 & 22.61 \\
3970 & H\,I & 15.70 & 15.96 & 15.50 & 15.95 & 14.70 & 15.97 \\
4069 & [S\,II] & 2.15 & 1.94 & 1.83 & 1.47 & ... & ... \\
4076 & [S\,II] & 0.87 & 0.62 & ... & 0.46 & ... &  ...\\
4101 & H\,I & 26.80 & 25.90 & 25.80 & 25.86 & 24.00 & 26.00 \\
4340 & H\,I & 47.20 & 47.00 & 46.30 & 47.04 & 46.60 & 46.99 \\
4363 & [O\,III] & 5.26 & 7.75 & 12.72 & 12.86 & 4.99 & 7.22 \\
4416 & O\,II & 0.03 & 0.01 & ... & ... & ... &  ...\\
4471 & He\,I & 5.60 & 5.68 & 5.05 & 5.02 & 5.47 & 6.19 \\
4562,71 & Mg\,I] & 0.66 & 0.66 & 0.26 & 0.40 & ... & ... \\
4686 & He\,II & 0.02 & 0.02 & 2.02 & 2.01 & 19.70 & 21.52 \\
4711 & [Ar\,IV] & 0.40 & 0.78 & 2.12 & 2.30 & 2.71 & 2.69 \\
4740 & [A\,IV] & 0.65 & 1.10 & 1.84 & 1.80 & 2.28 & 2.35 \\
4861 & H\,I & 100.0 & 100.0 & 100.0 & 100.0 & 100.0 & 100.0 \\
4959 & [O\,III] & 337.0 & 272.8 & 302.0 & 341.5 & 269.0 & 275.1 \\
5007 & [O\,III] & 994.0 & 788.4 & 913.0 & 987.2 & 804.0 & 795.1 \\
5016 & He\,I & 2.79 & 3.34 & 1.96 & 3.01 & 1.86 & 3.62 \\
5158 & [Fe\,II] & 0.03 & 0.03 & ... & ... & ... & ... \\
5199 & [N\,I] & 0.51 & 0.98 & 0.54 & 2.22 & ... & ... \\
5411 & He\,II & ... & 0.02 & 0.17 & 0.27 & 2.97 & 1.67 \\
5517 & [Cl\,III] & 0.30 & 0.34 & 0.65 & 0.68 & 0.75 & 0.65 \\
5537 & [Cl III] & 0.53 & 0.60 & 0.56 & 0.55 & 0.64 & 0.66 \\
5755 & [N\,II] & 1.64 & 1.70 & 1.74 & 1.79 & 0.34 & 0.17 \\
5876 & He\,I & 15.84 & 15.12 & 14.06 & 13.21 & 17.77 & 16.50 \\
6300 & [O\,I] & 5.35 & 5.11 & 4.11 & 3.66 & ... & ... \\
6312 & [S\,III] & 1.23 & 2.56 & 1.92 & 4.93 & 0.69 & 1.45 \\
6363 & [O\,I] & 1.83 & 1.64 & 1.43 & 1.17 & ... & ... \\
6548 & [N\,II] & 22.40 & 24.25 & 28.00 & 27.25 & 2.69 & 3.01 \\
6563 & H\,I & 288.0 & 284.0 & 281.0 & 284.3 & 284.0 & 283.8 \\
6584 & [N\,II] & 70.40 & 71.34 & 81.30 & 80.16 & 8.33 & 8.84 \\
6678 & He\,I & 4.23 & 4.29 & 4.86 & 3.74 & 4.81 & 4.68 \\
6717 & [S\,II] & 2.03 & 2.40 & 8.86 & 8.95 & 1.06 & 1.05 \\
6731 & [S\,II] & 4.06 & 4.60 & 10.19 & 10.02 & 1.59 & 1.55 \\
7136 & [Ar\,III] & 16.30 & 11.55 & 14.30 & 14.42 & 13.32 & 8.13 \\
7319 & [O\,II] & 5.89 & 8.05 & 2.99 & 3.33 & 2.29 & 1.50 \\
7329 & [O\,II] & 4.42 & 6.54 & 2.39 & 2.68 & 1.80 & 1.21 \\
7751 & [Ar\,III] & 3.76 & 2.77 & 3.45 & 3.45 & ... & ... \\
8045 & [Cl\,IV] & 0.09 & 0.03 & 0.43 & 0.41 & 0.61 & 1.10 \\
8579 & [Cl\,II] & 0.16 & 0.15 & 0.14 & 0.12 & ... & ... \\
8617 & [Fe\,II] & 0.03 & 0.06 & ... & ... & ... &  ...\\
8665 & H\,I & 0.86 & 0.83 & 0.84 & 0.83 & 0.90 & 0.83 \\
8750 & H\,I & 1.08 & 1.05 & 1.06 & 1.05 & 1.13 & 1.05 \\
8862 & H\,I & 1.40 & 1.37 & ... & 1.37 & 1.46 & 1.37 \\
\hline
	\end{tabular}}
\end{table}

\subsection{PB\,4}

\begin{figure}
\centering
	\includegraphics[scale=0.50]{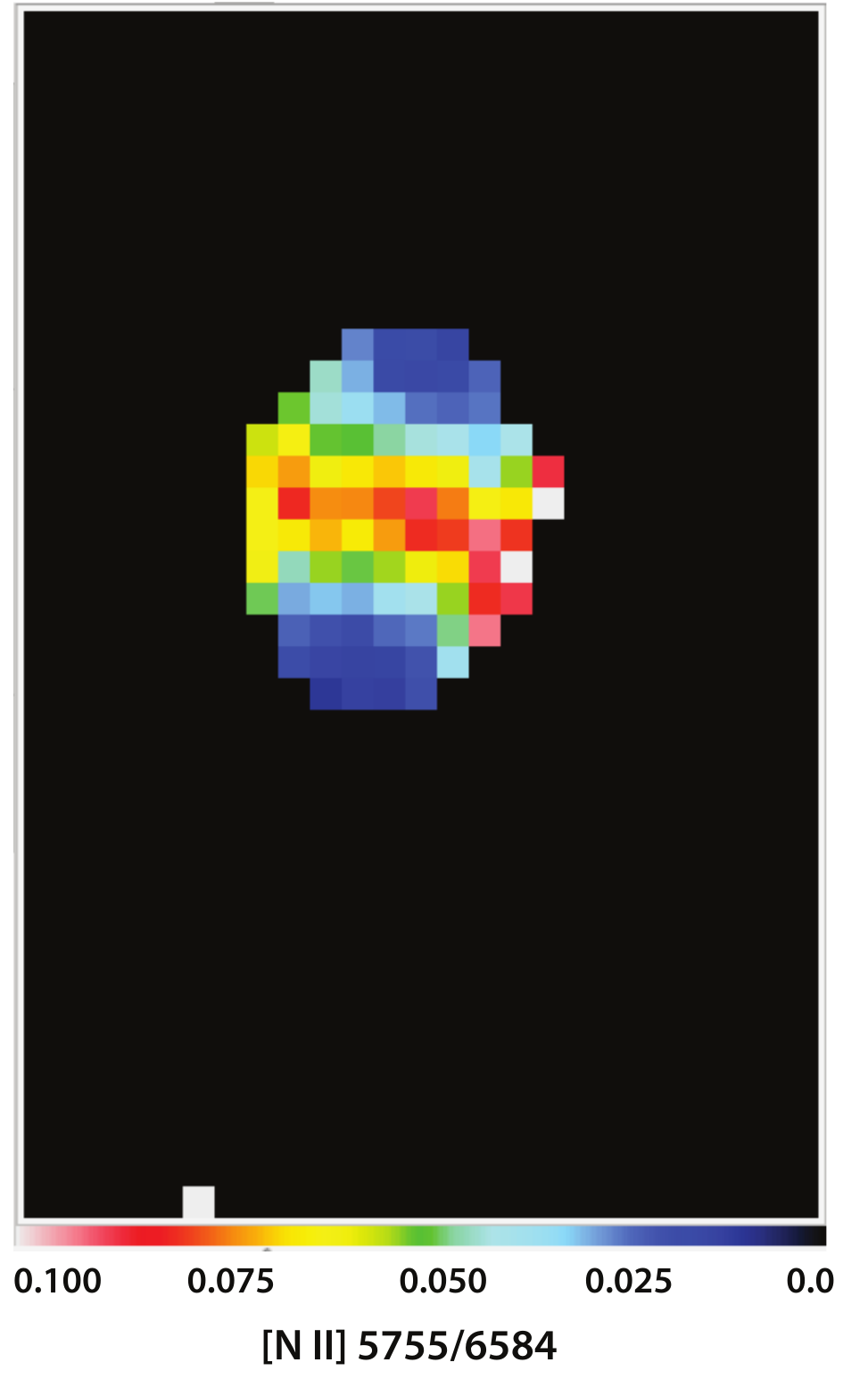}
	\caption{The spatial variation of the [N\,II]$\lambda\lambda5754.6/6583.5$ ratio in PB4. The polar axis is characterised by very high electron temperatures, possibly characteristic of a high ionisation parameter, or shock heating by a stellar wind.} \label{PB4-NII-ratio}
\end{figure}

This nebula is quite optically-thin, especially in the polar direction to the main elliptical ring (in which most of the flux is concentrated). However, an inspection of Figure \ref{Figure3} reveals a surprising anomaly. While the [O\,III] image is very similar to the H$\gamma$ image in its morphology, and [N\,II], as expected is confined to the elliptical ring, the temperature-sensitive [O\,II] $\lambda 7219.99$ line image brings out the polar extension of the nebula, seen clearly in the high-excitation lines, such as [Ar\,IV] $\lambda7237.26$ and in the O\,II $\lambda$4649.13 transition. On the other hand, the image in the [O\,II] doublet $\lambda\lambda3727.0,3728.8$, although of very poor quality due to the heavy reddening, shows a morphology resembling that of [N\,II]$\lambda6583.5$. This suggests that the plasma in the polar axis ($PA\sim 160^o$) is of much high temperature than that in the elliptical ring.

To investigate this possibility further, we constructed a line ratio image, as described above, in the [N\,II]$\lambda\lambda5754.6/6583.5$ ratio. This is shown in Figure \ref{PB4-NII-ratio}. It is clear that this temperature-sensitive [N\,II] ratio is strongly enhanced along the polar axis. There are two possible explanations, either that the plasma has an enhanced NII temperature in the polar cone, or that the plasma emitting the [N\,II]$\lambda5754.6$ has an extremely high density such that the [N\,II]$\lambda 6583.5$ and the [O\,II]$\lambda 3727,9$ lines are collisionally de-excited. At the measured electron density of the nebula  ($n_e \sim 2000$) the observed [N\,II]$\lambda \lambda 5754.6/6583.5$ line ratio $\sim0.085$ would imply an electron temperature of 30-40\,kK for the polar regions. If on the contrary  the electron temperature is $\sim 10$\,kK, then the electron density would need to be in the vicinity of $n_e \sim 10^5$cm$^{-3}$. Unlike the  O\,II $\lambda$4649.13 transition, the [N\,II]$\lambda5754.6$ line cannot be pumped by UV fluorescence directly from the ground state, so Figure \ref{PB4-NII-ratio} is clear proof of an enhanced N\,II temperature in a polar cone. Further, we subtracted a spectrum from the polar region to show the recombination contribution of N$^{+2}$ in the strength of the [N\,II]$\lambda5754$ line. The results show an enhancement of $\sim 70\%$ in the strength of the [N\,II]$\lambda5754$ line. This value is resemble that deduced from the spectrum of the entire nebula (Section 3.3).

Again, we used the procedure described in  \citet{Basurah16} by varying the assumed distance  to obtain a simultaneous fit for the nebula radius in the two zones of the model, as well as the absolute H$\beta$ flux. However, in this case, we required a very small optical depth in the H-ionising continuum to fit the polar region ($\tau = 1.3$), and we also require that the elliptical disk is somewhat optically-thin to the escape of H-ionising photons ($\tau =10$). This procedure gave a distance in good agreement with that given in Table \ref{Table3}; $3.1\pm0.3$\,kpc, giving an outer radius of $2.7\times10^{17}$\,cm for the inner ellipse (\emph{vs.} $2.4\times10^{17}$\,cm given by the model), and $1.9\times10^{17}$\,cm for the butterfly wings, \emph{vs.} $1.9\times10^{17}$\,cm as predicted by  the model. At this distance, the absolute H$\beta$ flux is $1.3\times10^{34}$\,erg/s \emph{vs.} $1.4\times10^{34}$\,erg/s for the model. The implied stellar luminosity of the central star is $L_* = (5\pm1.5)\times10^{36}$\,erg/s, or $1300\pm390 L_{\odot}$, and the stellar temperature $T_{\mathrm{eff}} =93\pm3.0$\,kK. The central disc contributes $\sim 57$\% of the total flux.

The L1-norm for the fit is 0.098 dex. This fit is poorer than the other PNe because the strength of the temperature-sensitive lines of [O\,II], [N\,II] are under-estimated by the model, while the [O\,III]$\lambda4363$ line is over-estimated. This points to the extra source of heating in the polar direction being due to stellar-wind driven shocks, rather than photoionisation acting alone. This is in addition to the evidence of EUV fluorescence derived from the morphology in the O\,II $\lambda$4649.13 transition.

An alternative explanation for these discrepancies, and for some of the recombination line abundance anomaly in this object could be that the electron distribution is a $\kappa-$distribution rather than a Classical Maxwell-Boltzmann distribution in the polar regions of this nebula, as a consequence of mechanical energy transport in the outflow \citep{Nicholls12}. Recently, \citet{Livadiotis18} has proved that  $\kappa-$distributions are the most general, physically meaningful, distribution function that particle systems are stabilised into when reaching thermal equilibrium, and that the Classical Maxwell-Boltzmann distribution simply represents a limiting case. As shown by \citet{Nicholls12}, the high energy tail of a  $\kappa-$distribution can enhance the temperature-sensitive collisionally-excited lines, while at the same time the excess of low-energy electrons in the $\kappa-$distribution enhances the recombination lines.

\begin{figure*}
\centering
	\includegraphics[scale=0.6]{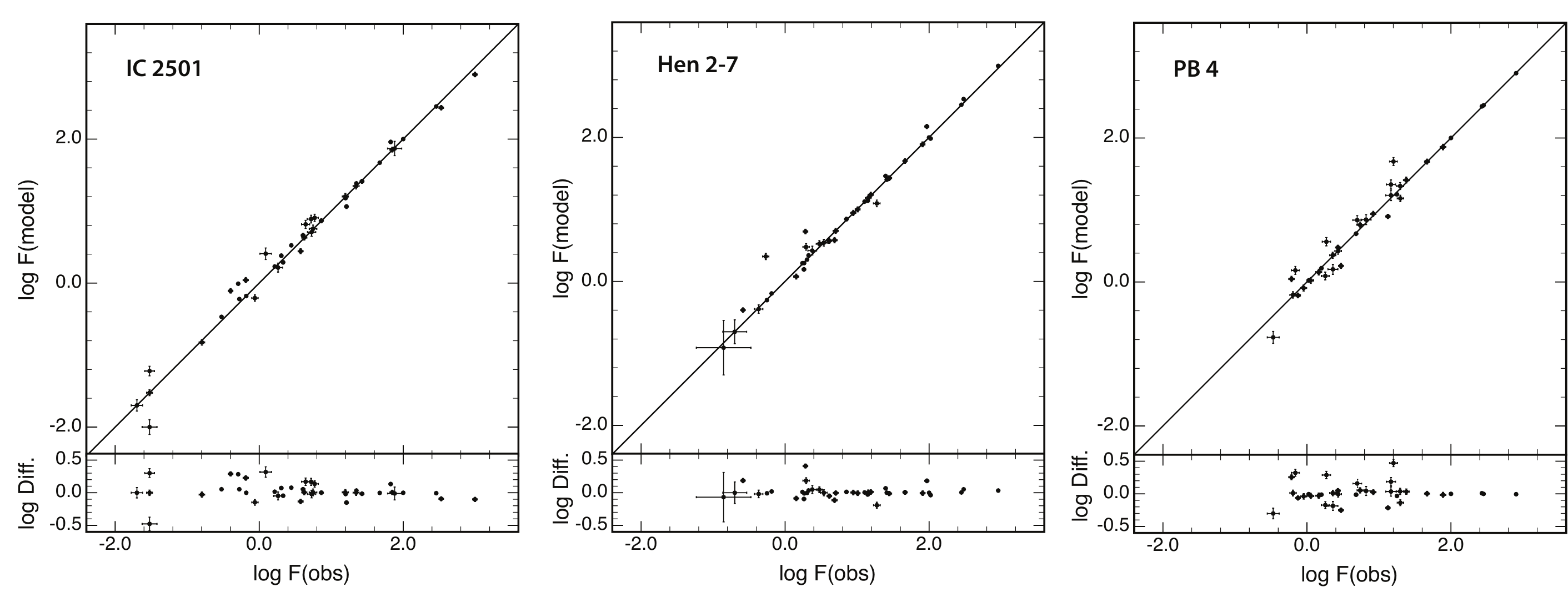}
	\caption{(Upper panels: The theoretical model predicted line fluxes plotted against the observed fluxes on a logarithmic scale. Lower panels: the logarithmic residuals of the model fit compared with the observations. The error bars show only the observational uncertainties.} \label{modelfit}
\end{figure*}

\subsection{Model fits to observations}
Overall the photoionisation models described above provide an excellent fit to the observations. In Figure \ref{modelfit} we show both the theoretical model fluxes and the residuals plotted against the observed line fluxes, all on a logarithmic scale. In Table \ref{Table9} we list the observed fluxes and model fluxes of the lines used in the fitting process. The full observed line list and their observational errors are given in Table \ref{Table2}. Overall the model fits are excellent. However, it is clear that the models systematically over-estimate [S\,III]$\lambda 6312$, and the [N\,I]$\lambda 5198,5200$ doublet for reasons that are obscure to us.

\begin{figure}
	\includegraphics[scale=0.65]{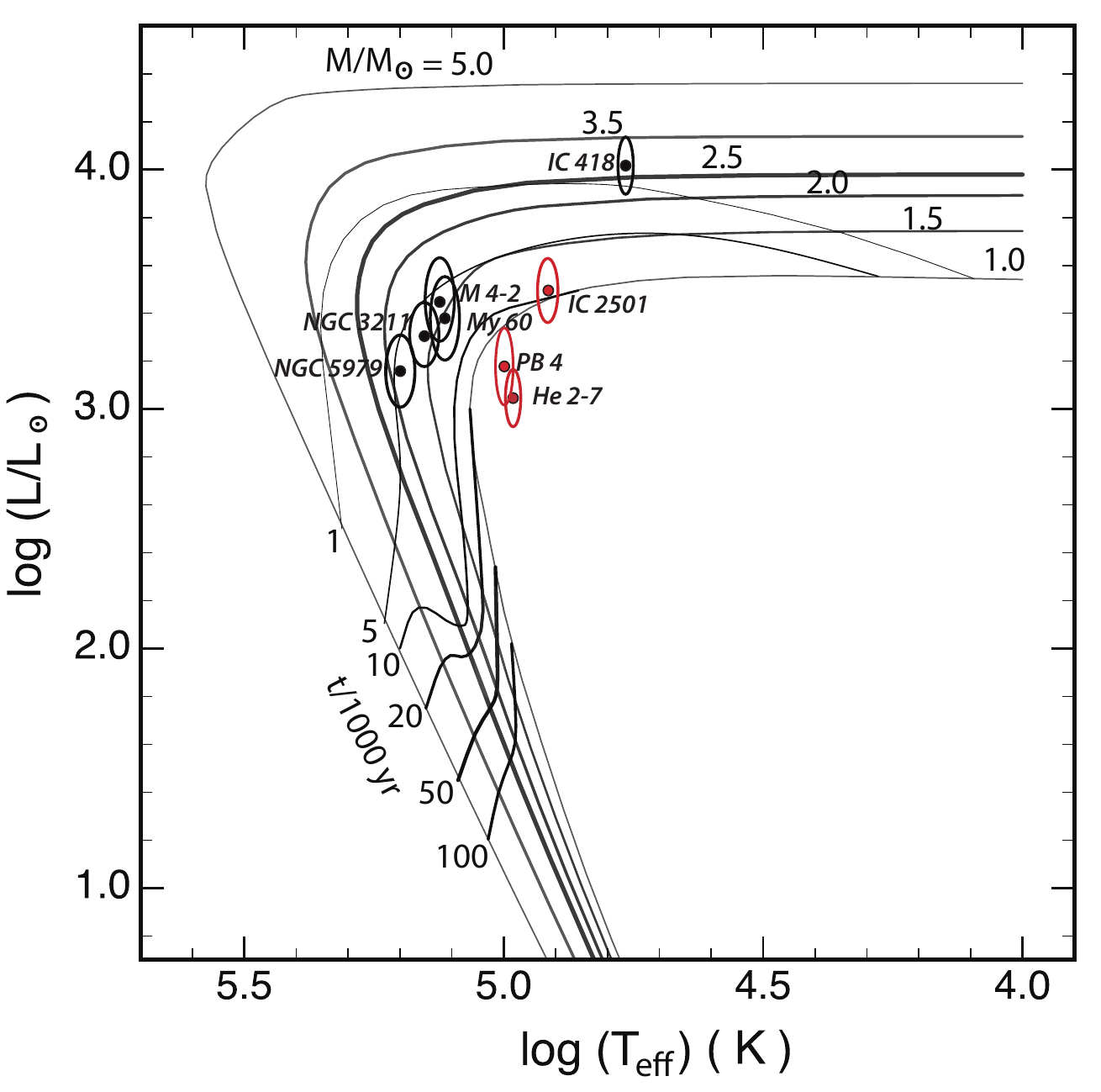}
	\caption{The H-R diagram for the PNe analysed so far, with their measurement error ellipses. The objects from this paper are marked in red. The Hydrogen-burning evolutionary tracks of the central stars of PNe are taken from \citet{VW94}, and are marked according to the initial stellar mass and the age of the PNe. Note that, assuming that the lie on Hydrogen-burning tracks, the implied mass for these objects is $M\sim 1.0M_{\odot}$. However, it is more likely that these objects are Helium-burning PNe of higher mass.} \label{HR-diag}
\end{figure}

\subsection{Inferred progenitor masses and nebular ages}
In Figure \ref{HR-diag}, we show the positions of these objects on the Hertzprung-Russell (H-R) Diagram compared with the tracks for Hydrogen-burning objects from \citet{VW94}, along with the objects which we previously analysed by these techniques (\citet{Basurah16},  \citet{Dopita17}). The implied mass for the three PNe studied here is $M\sim 1.0M_{\odot}$, or less, if these are interpreted as Hydrogen-burning objects. However, it is more likely that they represent higher-mass Helium-burners which are less luminous in this part of the H-R Diagram. If Helium-burning, then these objects have masses in the range  $1.0 \leqslant M/M_{\odot} \leqslant 1.5$. From their expansion velocities we deduce nebular ages of 3200\,yr (IC2501), 4500\,yr (He 2-7) and 4130\,yr (PB 4). These ages would be consistent with  $M\sim 1.5M_{\odot}$ Helium-burning nuclei \citep{VW94}.

\section{Conclusions}
In this paper, we have analysed three galactic PNe using high-resolution integral field spectroscopy, performed detailed photoionisation analyses using classical ICF techniques, and constructed two-zone photoionisation models to reproduce both the dimensions and the absolute H$\beta$ fluxes of these objects.
Although all three objects studied here are almost at the same evolutionary state as far as the PN is concerned, they are strikingly heterogeneous in term of their structure and morphology. One is almost spherical and dense, a second is an elongated  bipolar with a strong ionised jet in rapid expansion, and the third has a high-temperature polar conical region with strongly enhanced permitted lines in N\,II and O\,II. The calculated oxygen abundance from recombination lines for PB\, 4 nebula is  found  to  be  discrepant  by  a  factor  of  $\sim$ 18 relative to that calculated from collisionally excited lines. This result places PB\,4 nebula in the select class of PNe that display extreme abundance discrepancy factors, which \citet{Wesson18} has identified as possessing short-period interacting binary central stars. From the comparison of the nebula morphology in several ions we suggest that this abundance discrepancy factor is probably in major part due to EUV fluorescence in the O$^+$ ion, rather than as the result of recombination of O$^{2+}$ to O$^+$. In this case, the EUV radiation field from the presumed binary nucleus appears to be strongly beamed in the poleward direction.

Furthermore, both He2-7 and PB 4 show evidence of a strong polar-directed stellar wind. In the case of He 2-7, we can infer the velocity of outflow from the [Fe\,III] dynamics, using the ellipticity of the equatorial ring to determine the angle of inclination to the line of sight. This gives $v_w = 88\pm10$\,km/s.  In the case of PB 4, the very high electron temperature deduced from the [N\,II] lines (which also seems to apply to the [O\,II]) suggests the operation of shocks at the boundary layer of a polar-directed stellar wind. No such enhancement is seen in the [O\,III] lines, and additionally, we find no evidence of an enhanced expansion rate in the high-temperature plasma.

\section*{acknowledgements}
The authors would like to thank the anonymous referee for drawing attention to the extreme abundance discrepancies in the PB4 nebula and also for valuable and constructive comments which have greatly improved the manuscript. MD acknowledges the support of the Australian Research Council (ARC) through Discovery project DP16010363. Parts of this research were conducted by the Australian Research Council Centre of Excellence for All Sky Astrophysics in 3 Dimensions (ASTRO 3D), through project number CE170100013.

\bibliographystyle{mn2e_new}
\bibliography{PaperVII_PNe}

\end{document}